\documentclass[fleqn,usenatbib]{mnras}

\usepackage[T1]{fontenc}
\usepackage{ae,aecompl}

\usepackage{graphicx}	% Including figure files
\usepackage{amsmath}	% Advanced maths commands
\usepackage{amssymb}	% Extra maths symbols
\usepackage{bm}

\newcommand{\expect}{\mathbb{E}}
\newcommand{\disp}{\mathbb{D}}
\newcommand{\const}{\mathop{\rm const}\nolimits}
\newcommand{\FAP}{{\rm FAP}}
\newcommand{\TAP}{{\rm TAP}}
\title[Frequentist vs. Bayesian periodic signal detection]{Comparing the
frequentist and Bayesian periodic signal detection: rates of statistical mistakes and
sensitivity to priors}

\author[R.V.~Baluev]{Roman V. Baluev\thanks{E-mail: r.baluev@spbu.ru}\\
Saint Petersburg State University, 7--9 Universitetskaya Emb., Saint Petersburg 199034, Russia}

\begin{document}

\date{Accepted 2022 March 16.
      Received 2022 March 16;
      in original form 2022 February 18}

\pagerange{\pageref{firstpage}--\pageref{lastpage}} \pubyear{2022}

\maketitle

\label{firstpage}

\begin{abstract}
We perform extensive Monte Carlo simulations to systematically compare the frequentist and
Bayesian treatments of the Lomb--Scargle periodogram. The goal is to investigate whether
the Bayesian period search is advantageous over the frequentist one in terms of the
detection efficiency, how much if yes, and how sensitive it is regarding the choice of the
priors, in particular in case of a misspecified prior (whenever the adopted prior does not
match the actual distribution of physical objects). We find that the Bayesian and frequentist
analyses always offer nearly identical detection efficiency in terms of their tradeoff
between type-I and type-II mistakes. Bayesian detection may reveal a formal advantage if
the frequency prior is nonuniform, but this results in only $\sim 1$ per cent extra
detected signals. In case if the prior was misspecified (adopting nonuniform one over the
actual uniform) this may turn into an opposite advantage of the frequentist analysis.
Finally, we revealed that Bayes factor of this task appears rather overconservative if used
without a calibration against type-I mistakes (false positives), thereby necessitating such
a calibration in practice.
\end{abstract}


\section{Introduction}
Periodogram is a tool that enables the detection of periodic variations in time series
data, a widespread astronomical task. Speaking more accurately, it is a collection of
tools, because multiple periodograms were constructed so far. The earliest one is the
classic periodogram, also known as the Schuster periodogram \citep{Schuster1898}. It
originates from a direct estimation of the power spectrum.

However, another type of periodograms became more widespread in astronomy, known as the
\citet{Lomb76} -- \citet{Scargle82} periodogram (LSP hereafter). It is a representative of
an alternative approach based on the least squares spectral analysis (LSSA), or the
Van{\'\i}{\v{c}}ek method \citep{Vanicek69}. Unlike the Schuster periodogram, the LSP was
not defined through the power spectrum (even though it can be used as such an estimate if
necessary). Rather, it was defined through the least squares fit of the data using a
sinusoidal model.

One reason why the LSP appeared so popular in astronomical applications (perhaps contrary
to the Schuster periodogram) is flexibility of the LSSA behind it. Periodograms of this
type can be easily generalized to include more complicated models. In particular,
\citet{FerrazMello81} introduced the so-called date-compensated discrete Fourier transform
(DCDFT) that involves an internally fittable constant offset (not present in the initial
LSP). Now it is also known as the generalized LSP (or GLSP), thanks to the efficient
computing algorithm by \citet{ZechKur09}. \citet{Cumming99} considered more extensions of
this type taking into account linear or polynomial trends. Further ways to generalize the
LSP involve the use of nonsinusoidal and/or nonlinear models of the periodic signal,
multicomponent signals, as well as more complicated noise models, including even Gaussian
processes, see \citep{Baluev14c} for a review and
\citep{Baluev08b,Baluev09a,Baluev13a,Baluev13d,Baluev14b} as particular examples.

All these generalizations of the LSP refer to the frequentist analysis. A yet another
branch of period search tools is based on the Bayesian treatment of a periodogram. There are
two main ways how the Bayesian periodograms were introduced in the literature. The first one is
due to \citet{Bretthorst01a,Bretthorst01b}, who defined the Bayesian LS periodogram (or BLSP)
through the posterior frequency distribution. This approach was further developed by
\citet{Mortier14} who added a free offset to the model, just like in the DCDFT/GLSP, so
their method is called as the Bayesian GLSP (BGLSP).

The second way to construct a Bayesian periodogram follows from \citet{Cumming04}, and it
is based on the Bayes factor (or a close entity like the posterior odds ratio) of the
corresponding signal detection task, rather than on the posterior frequency distribution.

Two flavors of a Bayesian periodogram mentioned above are not necessarily equivalent,
because the Bayes factor is deemed to solve the signal detection task (basically,
determination of an absolute ordinate scale for a periodogram), while the posterior
frequency distribution targets the aliasing ambiguity task (relative intercomparison of
peaks in the same periodogram). In this study we focus on the signal detection, which is
more basic with respect to the aliasing (before resolving possibly ambiguous frequency, we
should demonstrate that some periodic signal is likely to exist). Therefore, we assume from
the beginning that our Bayesian periodogram is based on the Bayes factor.

Bayesian periodograms, just like Bayesian analysis in general, are usually believed to
offer some advantages over the classic frequentist ones. In particular, they offer an easy
way to take into account prior information, if it is available in the form of a more or
less accurately defined distribution. From the other side, they have an obvious practical
disadvantage because of high computing demands. Computation of a frequentist periodogram
may be slow sometimes, maybe not too much faster than Markov Chain Monte Carlo (MCMC)
typically used to sample a posterior distribution. However, Bayes factor computation is
more challenging than MCMC. Therefore, an important practical question is raised: do
the Bayesian periodograms actually offer an advantage worthy of the sacrificed computing speed?
For example, do they offer a better detection efficiency, and if yes then how valuable this
advantage is? And is it stable with respect to input conditions like time series sampling,
amount of data, adopted priors, etc? These questions are often avoided in practice, because
different works usually tend to apply a particular selected method to a particular
practical task. However, to gain a more or less completed picture it is necessary to test
two rival approaches systematically, under varying conditions, but always using an
equivalent treatment for them both. Here our aim is to perform such a comparison, based on
numerical simulations.

One recent example of such a study, employing a good-practice comparison, was provided by
the exoplanets detection challenge by \citet{Nelson20}. However, this work was focused,
mostly, on the intercomparison of different Bayesian methods, mainly considering the
stability of the computed Bayesian evidence. The detection efficiency for Bayesian and
frequentist methods was not considered in an equivalent and comparative treatment.
Additionally, this challenge targeted more complicated data models developed for a more
specific task of exoplanets detection. Our goal here is to consider a wider-purpose
formulation of the period search task.

The plan of the paper is as follows. In Sect.~\ref{sec_bayes} we present an introductory
discussion about the general relationship between Bayesian and frequentist analyses, how
they may be formally compared, and how this applies to periodograms. In
Sect.~\ref{sec_layout} we present the details of our Monte Carlo simulations, while the
results of these simulations are presented in Sect.~\ref{sec_results}. Most of these
results are presented in graphical form, but in the printed paper we provide only a portion
of the plots for demonstrative purposes. The full set of graphs is attached as the
online-only material (see App.~\ref{sec_OO}).

\section{Bayesian vs. frequentist treatments of the periodogram}
\label{sec_bayes}
\subsection{Relationship between Bayesian and frequentist model checking}
Bayesian and frequentist analysis differ from each other at multiple layers, starting from
the philosophical understanding of probability \citep{Gelman08}. Of course, a complete
review of the issue is out of the scope here, because our main goal is to investigate a
particular task of periodic signal detection. Still there are practical aspects that we
need to compare in a more general context to avoid certain misconceptions.

In Bayesian model selection we deal with the so-called evidence:
\begin{equation}
E_{\rm B}(\bmath x | \mathcal M, \mathcal W) = \int L(\bmath x|\btheta,\mathcal M) \mathcal W(\btheta) d\btheta,
\label{evB}
\end{equation}
where $L$ is the likelihood, i.e. the p.d.f. of the data $\bmath x$ given the model
$\mathcal M$ and its parameters $\btheta$. The weight function $\mathcal W(\btheta)\geq 0$
is the prior p.d.f. for $\btheta$.

The evidence $E_B$ is also known under other titles: prior predictive, average likelihood,
marginal likelihood. In fact, it represents the p.d.f. of the data $\bmath x$, treated
conditionally to the adopted model and to the adopted prior $\mathcal W$.

The ratio of two Bayesian evidences like~(\ref{evB}), computed for two alternative models
is called the Bayes factor
\begin{equation}
R_{\rm B}(\bmath x) = \frac{E_{\rm B}(\bmath x | \mathcal M_2, \mathcal W_2)}{E_{\rm B}(\bmath x | \mathcal M_1, \mathcal W_1)}.
\label{BF}
\end{equation}
It usually serves as a key quantity for Bayesian model selection \citep[e.g.][]{Bayes18}.

In the frequentist hypothesis testing we usually deal with ratios of likelihood maxima that
refer to our rival models. In this case we should first define the quantity
\begin{equation}
E_{\rm F}(\bmath x | \mathcal M, \Theta) = \max_{\btheta \in \Theta} L(\bmath x|\btheta,\mathcal M),
\label{evF}
\end{equation}
where $\Theta$ is some \emph{a priori} specified domain, in which $\btheta$ is supposed to
always reside. The maximum likelihood, $E_{\rm F}$, plays now the same key role as the
evidence $E_{\rm B}$ served in Bayesian analysis. Therefore, we may alternatively call it
the ``frequentist evidence''. And analogously to Bayesian case, we can define the ratio of
two rival evidences which now turns into the classic likelihood ratio:
\begin{equation}
R_{\rm F}(\bmath x) = \frac{E_{\rm F}(\bmath x | \mathcal M_2, \Theta_2)}{E_{\rm F}(\bmath x | \mathcal M_1, \Theta_1)}.
\end{equation}
We can see that $R_{\rm F}$ apparently disallows to include any detailed prior distribution
(in form of a weight function). This is a point the frequentist approach is sometimes
criticized for, because $R_{\rm B}$ just looks more versatile and offers more control
through $\mathcal W_{1,2}$. But simultaneously it is neither true that frequentist approach
ignores any prior information at all (there is a prior domain $\Theta$), nor that it
assumes any specific prior in Bayesian sense (because $E_{\rm F}$ is not simply a
restriction of $E_{\rm B}$ to any specific prior).

From a practical point of view, Bayesian and frequentist approaches are constructively
similar, but differ in their target metric, either $E_{\rm B}$ or $E_{\rm F}$. The first
one is based on averaging, while the second one relies on optimization. But let us now
consider the following ``generalized evidence'':
\begin{equation}
E_p(\bmath x | \mathcal M, \mathcal W) = \left(\int \left[L(\bmath x|\btheta,\mathcal M)\right]^p \mathcal W(\btheta) d\btheta \right)^{\frac{1}{p}}.
\label{evLp}
\end{equation}
This is simply the $L_p$-norm of the likelihood. Now we can see that both $E_{\rm B}$ and
$E_{\rm F}$ represent two limiting cases of $E_p$:
\begin{eqnarray}
E_{\rm B}(\bmath x | \mathcal M, \mathcal W) &=& E_1(\bmath x|\mathcal M, \mathcal W),
\nonumber\\ E_{\rm F}(\bmath x | \mathcal M, \Theta) &=& E_\infty(\bmath x | \mathcal M, \mathcal W), \quad \Theta = {\rm supp}(\mathcal W).
\end{eqnarray}
And based on the generalized $L_p$-evidence~(\ref{evLp}), the Bayes factor~(\ref{BF}) is
generalized to
\begin{equation}
R_p(\bmath x) = \frac{E_p(\bmath x | \mathcal M_2, \mathcal W_2)}{E_p(\bmath x | \mathcal M_1, \mathcal W_1)},
\end{equation}
so that $R_{\rm B}=R_1$ and $R_{\rm F}=R_\infty$.

From this position Bayesian and frequentist approaches look entirely as peers. None of the
two metric is a worse (or weakened) version of another; rather they are opposite special
cases of the same $E_p$. They differ from each other just like the $L_1$-norm differs from
the $L_\infty$-norm. This argumentation also helps us to understand better why the $E_{\rm
F}$ metric does not apparently allow to specialize any detailed prior. Before passing to
the limit $p\to\infty$, the evidence does include $\mathcal W$ in its full form, but in the
limiting behaviour it appears \emph{invariable} with respect to it. Setting different
$\mathcal W$ results in different $E_{\rm B}$, but always leads to the same $E_{\rm F}$, at
least if $\Theta$ is unchanged.

If the ``generalized evidence'' $E_p$ looks too artificial, it could be constructed using
an alternative more constructive argumentation. Bayesian evidence~(\ref{evB}) depends on
the prior $\mathcal W(\btheta)$ in the form of a penalty added to the log-likelihood, i.e.
as $\log L + \log \mathcal W$. However, in an observations-related science nothing can be
specified precisely, and so $\mathcal W$ may involve uncertainties. In this case we should
modify the evidence to take the uncertainty of $\mathcal W$ into account. Basically, we
should downweight the impact of $\mathcal W$, for example as $\log L +
\frac{1}{p} \log \mathcal W$, where $p\geq 1$. And precisely this type of downweighting is
offered by the definition~(\ref{evLp}).\footnote{It is also important that $\mathcal W$
retains the role of an integration measure, since its physical meaning is probability. This
is why we cannot introduce something like e.g. $\int L {\mathcal W}^{1/p} d\btheta$.} The
quantity $p$ attains now the meaning of the prior's uncertainty. The minimum convex metric
has $p=1$, and it corresponds to the pure Bayesian analysis with a strictly specified
prior. The frequentist analysis, on the other hand, corresponds to the maximum-uncertainty
case with infinite $p$.

Of course, if prior information is sound enough, e.g. if we can formalize our physical
knowledge through a well-determined function $\mathcal W$, then it looks reasonable to use
Bayesian approach, because it can be tuned against this particular $\mathcal W$. But we
often cannot formalize our knowledge through a well-determined function. In this case, a
common Bayesian workaround is to use some non-informative (high-entropy) prior, e.g.
wide-range flat one. However, such $\mathcal W$ is just a yet another strictly specified
prior. It may appear wrong, that is it may mismatch the actual distribution of physical
objects that we consider. Consequences of such a mismatch are unclear and need an
investigation, at least.

For example, in the task of periodic signal detection it looks physically reasonable to
assume a uniform prior for an angular variable like the phase of a sinusoid. Contrary to
that, the frequency prior is less obvious, and in fact applications may assume different
frequency priors. For example, it can be uniform in terms of the frequency $f$ itself, or
in terms of the period $P=1/f$, or in terms of $\log f$. These are three quite different
and rather restrictive distributions that may or may not match the distribution of actual
objects that we observe. Similar uncertainty appears regarding the prior for the amplitude
of a periodic signal.

Summarizing all the above, three primary questions can be formulated: (i) what happens if
our adopted Bayesian prior $\mathcal W$ mismatches the actual distribution of physical
objects, (ii) whether and when the Bayes factor $R_{\rm B}$ becomes advantageous or
disadvantageous over the frequentist likelihood ratio $R_{\rm F}$, and (iii) how to measure
this possible advantage numerically. However, in this study these questions are to be
addressed in view of our practical task, the detection of a periodic signal. In this
setting the $\mathcal M_1$ model refers to noise + possible nuisance background variation,
while $\mathcal M_2$ additionally contains a periodic signal.

\subsection{Statistical error rates: why do we need them?}
Differences between Bayesian and frequentist treatments are not limited to the use of
different statistic, $R_{\rm B}$ or $R_{\rm F}$. The further difference is how these $R$
values are used.

Let us consider the frequentist case first. In general, larger values of the likelihood
ratio $R_{\rm F}$ indicate better support in favor of the model $\mathcal M_2$, over
$\mathcal M_1$. However, the practical numeric threshold of $R_{\rm F}$ is derived from
certain post-calibration. This calibration is usually done through the notion of $p$-value,
aka false alarm probability (more frequent term in astronomy), aka false positives rate,
aka type-I mistake probability. The false alarm probability ($\FAP$) is the probability to
obtain, being governed by the threshold rule $R_{\rm F}\geq R_{\rm F}^*$, an erroneous
decision in favour of $\mathcal M_2$ while $\mathcal M_1$ is actually true. Whenever this
$\FAP$ was expressed through a function $\alpha(R_{\rm F}^*)$, and given some small level
$\alpha^*$, we may define the threshold $R_{\rm F}^*$ from the implicit equation
$\alpha(R_{\rm F}^*) = \alpha^*$.

While the type-I mistakes rate provides the required calibration for $R_{\rm F}$, type-II
mistakes define another important characteristic, namely the power of the test, or how well
it can reveal hidden low-amplitude signals. Given the threshold $R_{\rm F}^*$, the type-II
mistake is a false non-detection of an existing signal, meaning to accept $\mathcal M_1$
while $\mathcal M_2$ is true. The type-II mistakes rate is characterized by another
function $\beta(R_{\rm F}^*)$.

\begin{figure*}
\includegraphics[width=\linewidth]{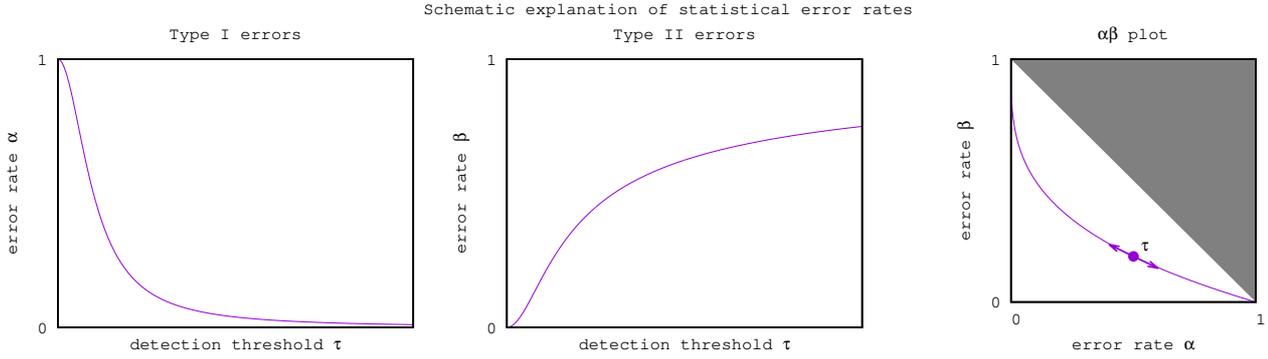}
\caption{Illustration explaining the $\alpha$ and $\beta$ error rates and the tradeoff
between them. The argument $\tau$ stands for the adopted test statistic, either $R_{\rm F}$
or $R_{\rm B}$ or whatever. Notice that $1-\beta$ has the meaning of ``true alarm
probability'', which should necessarily exceed $\alpha$, at least for a logically
reasonable test. So in the $\alpha\beta$ diagram we grayed out the top-right triangle
$\alpha+\beta>1$ to reflect this.}
\label{fig_err}
\end{figure*}

Therefore, we ultimately have two univariate functions that jointly characterize error
rates of our test: $\alpha(R_{\rm F})$ and $\beta(R_{\rm F})$. The $\alpha$-function is
monotonically decreasing, while $\beta$ is monotonically increasing, so the choice of the
detection threshold $R_{\rm F}^*$ is always governed by the tradeoff between $\alpha$ and
$\beta$. In fact, the $R_{\rm F}$ argument becomes now just a dummy variable, so its
calibration (absolute scale) is not important in itself. What attains the greatest
importance is the implicit relationship between $\alpha$ and $\beta$. 2D diagrams of this
type allow us to comprehensively characterize the intrinsic test performance, as shown in
Fig.~\ref{fig_err}. They will be our primary tool to investigate the performance of a
signal detection test.

Now let us proceed to Bayesian signal detection. There is a universal empiric scale to
interpret $R_{\rm B}$ \citep{JeffreysTP}. For example, $R_{\rm B}=1$ means equally
supported models, $R_{\rm B}=10$ means substantial evidence in favour of $M_2$, $R_{\rm
B}=100$ means very strong evidence for $M_2$, and so on. Contrary to the frequentist case,
we may finish our analysis by computing $R_{\rm B}$, apparently without any need of its
calibration. The associated rates of statistical mistakes are then moved out of the focus
entirely, and that is why they are often deemed to be a solely frequentist-related tool.

However, even in Bayesian literature it is recognized that without a calibration against
the mistakes rates any statistical method (including Bayesian) may appear deficient
\citep{Bayes20}. Any statistical method must be calibrated in the sense that we should have
an assessment of how frequently it yields the right answer. And in fact, there is no
universal guarantee that Jeffreys' empiric scale for the Bayes factor remains equally good
for any particular task.

\begin{figure*}
\includegraphics[width=\linewidth]{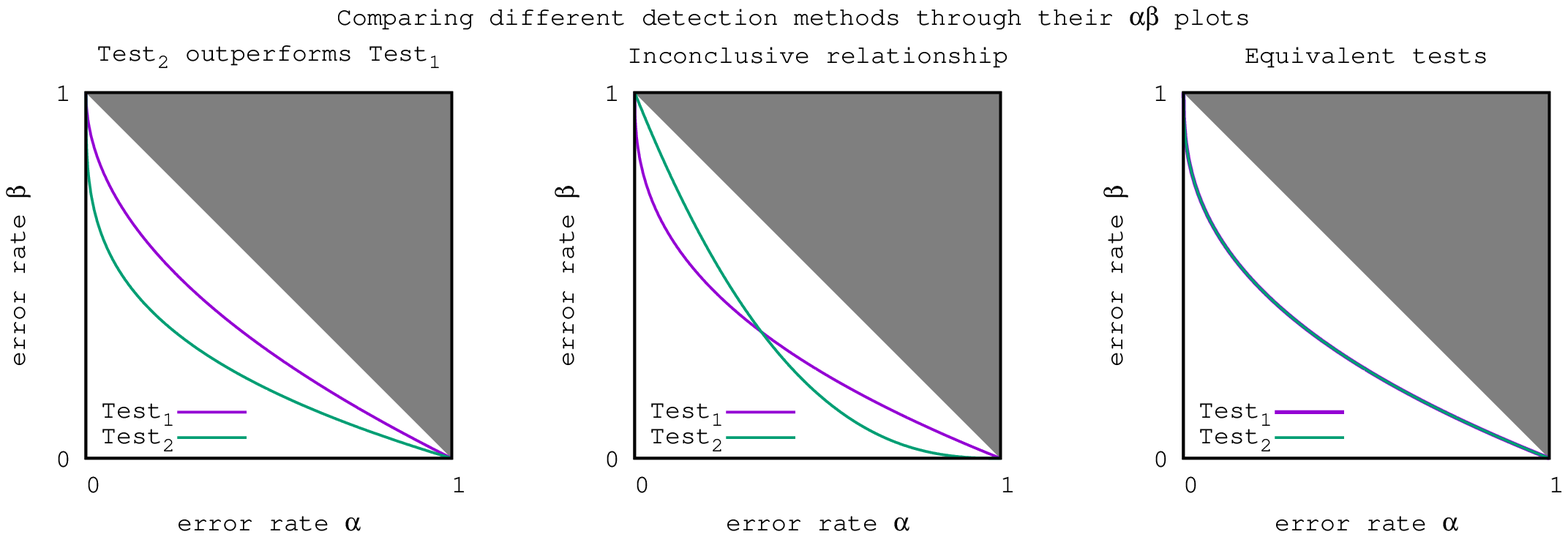}
\caption{Comparison of different tests using the $\alpha\beta$ diagram.}
\label{fig_errcomp}
\end{figure*}

In the setting of our study we should characterize the behaviour of $R_{\rm B}$ using the
$\alpha\beta$ diagram, entirely analogously to the frequentist likelihood ratio $R_{\rm
F}$. Moreover, the $\alpha\beta$ diagram appears useful to intercompare the performance of
such alternative statistical tests, see Fig.~\ref{fig_errcomp} for an illustration. To make
this comparison a bit more objective, we introduce the following metric:
\begin{eqnarray}
I_{\alpha\beta} &=& \int\limits_0^1 \beta\, d\alpha = - \int\limits_{\tau_{\min}}^{\tau_{\max}} \beta(\tau) \alpha'(\tau) d\tau =\nonumber\\
 &=& \int\limits_{\tau_{\min}}^{\tau_{\max}} \alpha(\tau) \beta'(\tau) d\tau = \int\limits_0^1 \alpha\, d\beta = I_{\beta\alpha},
\label{Iab}
\end{eqnarray}
where $\tau$ is our test statistic, either $R_{\rm B}$ or $R_{\rm F}$. Smaller
$I_{\alpha\beta}$ implies that we have a smaller integrated rate of type-II mistakes,
meaning a more efficient detection in average. So tests with smaller $I_{\alpha\beta}$ are
likely more advantageous. Equal values of $I_{\alpha\beta}$ may mean that two methods are
either equivalent (third plot of Fig.~\ref{fig_errcomp}) or their relationship is
inconclusive (second plot of Fig.~\ref{fig_errcomp}).

In this section we cannot bypass a significant practical issue related to the calculation
of the $\alpha\beta$ rates. The $\FAP$, for example, appears to additionally depend on the
$\mathcal M_1$ model parameters:
\begin{equation}
\FAP(\tau^* | \btheta_1) = \Pr\{ \tau(\bmath x) > \tau^* \,|\, \btheta_1, \mathcal M_1 \}.
\label{FAPdef}
\end{equation}
The nuisance vector $\btheta_1$ is a priori unknown, but we need to reduce this $\FAP$ to
an univariate function $\alpha(\tau^*)$ somehow. A widespread practical workaround of this
issue is to substitute the best-fitting estimate $\hat\btheta_1$ in~(\ref{FAPdef}). This
simplified method is called the ``plug-in $p$-value'' by
\citep{Bayes18}:
\begin{equation}
\alpha_{\rm plug}(\tau^*) = \FAP(\tau^* | \hat \btheta_1).
\label{alphaplug}
\end{equation}
This method is obviously flawed because the estimate $\hat\btheta_1$ always has some
uncertainty, so the $\FAP$ resulting from~(\ref{alphaplug}) becomes inherently inaccurate.
In fact it is a yet another point the frequentist approach is sometimes criticized for,
however we can see that this issue is unrelated to the choice of the frequentist or
Bayesian detection metric $R_{\rm F}/R_{\rm B}$ and can equally affect the both. Moreover,
multiple more neat methods are known well that allow to get rid of the $\btheta_1$
dependency.

The first approach is now called the supremum method \citep{BB00}, although it was known
decades ago \citep[e.g.][\S~20.5]{Koroluk}. It requires to \emph{maximize} the $\FAP$ over
$\btheta_1$:
\begin{equation}
\alpha_{\rm sup}(\tau^*) = \max_{\btheta_1\in\Theta_1} \FAP(\tau^* | \btheta_1).
\label{alphasup}
\end{equation}
This method is frequentist-leaned because it is more conservative and allows to keep the
number of false positives below the requested $\alpha$ level regardless of the unknown
$\btheta_1$. Therefore, it is a natural solution if placing such a limit on the type-I
mistakes is our ultimate goal. A yet another alternative is to \emph{average} the $\FAP$
over $\btheta_1$ using the relevant prior:
\begin{equation}
\alpha_{\rm avg}(\tau^*) = \int \FAP(\tau^* | \btheta_1) \mathcal W_1(\btheta_1) d\btheta_1.
\label{alphaavg}
\end{equation}
This method is also called the prior predictive p-value \citep{Box80} and it is
Bayesian-leaned in some sense. It allows to count the cumulative number of false positives
in our survey, provided that $\mathcal W_1(\btheta_1)$ describes its actual (physical)
distribution. This list of possible ways to handle nuisance parameters is incomplete, for
example we omitted the posterior predictive p-values.

The error rate $\beta$ is a complementary probability to the following True Alarm
Probability ($\TAP$):
\begin{equation}
\TAP(\tau^* | \btheta_2) = \Pr\{ \tau(\bmath x) > \tau^* \,|\, \btheta_2, \mathcal M_2 \},
\label{TAPdef}
\end{equation}
As we can see, the $\TAP$ is also affected by the issue of unknown parametric argument, now
$\btheta_2$. And by analogy with the $\FAP$, we can use three alternative methods to deal
with this issue: (i) the plug-in method $\beta_{\rm plug} = 1 - \TAP(\tau^*,
\hat\btheta_2)$, it is the easiest one to calculate but flawed due to the uncertainty, (ii)
the infinum or supremum methods $\beta_{{\rm inf}\atop {\rm sup}} = 1 - {\max\atop\min}
\TAP(\tau^*, \hat\btheta_2)$, and (iii) the average method $\beta_{\rm avg} = 1 - \int
\TAP(\tau^*, \hat\btheta_2) \mathcal W_2(\btheta_2) d\btheta_2$.

Notice that although we said ``frequentist-leaned'' and ``Bayesian-leaned'' above, it does
not mean that the choice of the calibration method is tied to the choice of the $R_{\rm F}$
or $R_{\rm B}$ statistic in place of $\tau$. Depending on our goals, all combinations are
plausible, even mixed ones with different choices for $\alpha$ and $\beta$. It is only
important that when comparing two alternative tests we should use identical calibration
scheme for the both.

The biggest practical issue here is to compute~(\ref{alphasup}) or~(\ref{alphaavg}), and
their $\beta$ analogues. Fortunately, regarding $\alpha$ all this appeared unnecessary in
the context of our study, because the LSP has an empty (parameterless) model $M_1$ (see
below). In this case the function $\FAP(\tau)$ is independent of $\btheta_1$ from the very
beginning, implying that $\alpha_{\rm plug} = \alpha_{\rm sup} = \alpha_{\rm avg}$. So only
the choice of the $\beta$ calibration is important for our study. We adopt the average
method for $\beta$, because we are interested in counting cumulative number of
nondetections in the survey.

Calibration of the frequentist likelihood ratio $R_{\rm F}$ against the $\FAP$ is an old
and long-studied problem. A lot of analytic results are available here, in particular the
LSP $\alpha(R_{\rm F})$ function was approximated in \citep{Baluev08a}. However,
calibrating Bayes factor against $\alpha\beta$ rates appears relatively uncommon, so it is
hard to find useful analytic results for that. We therefore intend to rely on Monte Carlo
simulations below. However, in Appendix~\ref{sec_BFAP} the following simple inequality is
derived:
\begin{equation}
R_{\rm B}^* \leq \frac{1-\beta_{\rm avg}}{\alpha_{\rm avg}} \leq \frac{1}{\alpha_{\rm avg}} \implies \alpha_{\rm avg} \leq \frac{1}{R_{\rm B}^*}.
\label{Rineq}
\end{equation}
Simulations revealed that this inequality is not of a big practical value perhaps, because
it appears very overconservative. Nevertheless, it may provide an important additional
validation of the results.

\subsection{Application to the Lomb-Scargle periodogram}
\subsubsection{Task layout}
Now let us formulate our signal detection task and its basic input conditions. First of
all, our data vector $\bmath x$ contains $N$ measurements $x_i$ taken at the times $t_i$,
either even or uneven in general. The value of each $x_i$ is expressed through the sum of
the model $\mu_{1,2}(t_i,\btheta_{1,2})$, as defined by $\mathcal M_{1,2}$, and of the
white Gaussian noise (WGN) with mean zero and known standard error $\sigma_i$.

In this work we limit out attention to the LSP-based signal detection and its Bayesian
extension. The LSP assumes that model $\mathcal M_1$ is literally parameterless,
$\btheta_1=\emptyset$ and $\mu_1\equiv 0$. In this case $\bmath x$ represents just the raw
WGN. Under the model $\mathcal M_2$, it additionally contains a sinusoidal variation $\mu_2
= K\cos(2\pi f t + \varphi)$ with fittable frequency $f$, amplitude $K$, and phase
$\varphi$, so $\btheta_2=\{f,K,\varphi\}$.

Of course, the parameterless model $\mathcal M_1$ is a simplification. In practice input
data have some offset at least, so they must be precentered before the LSP can be formally
applied, usually by subtracting the mean. But in case of uneven timings simple subtraction
of the mean may become inaccurate because of correlations that appear between sine/cosine
functions and a constant \citep{Cumming99}. The neat way to handle this issue is to use an
expanded model $\mathcal M_1$ like in the DCDFT/GLSP \citep{FerrazMello81,ZechKur09}. A
general formulation with an arbitrary linear regression for $\mathcal M_1$ is given in
\citep{Baluev08a}. Further issue appears if the noise uncertainties $\sigma_i$ are not
known precisely, and it is commonly known as periodogram normalization issue
\citep{SchwCzerny98b}. It can be treated through additional noise parameters in $\mathcal
M_1$ \citep{Baluev08b}, so this issue is a yet another example of an oversimplified
$\mathcal M_1$.

Improved periodograms like DCDFT/GLSP are clearly more preferred over the LSP if we deal
with practical data (e.g. non-centered or trend-imposed ones), but only because the LSP
would then be in \emph{out-of-task conditions}. General statistical properties of such
periodograms do not appear to differ much from those of the LSP \emph{whenever the latter
is used in an in-task way}. For example, a low-order polynomial trend in $\mathcal M_1$
usually have only a negligible effect on the frequentist $\alpha$ curve of a periodogram
\citep{Baluev08a}. In this work we do not aim to analyse any practical data, since we need
to reveal only general tendencies regarding periodograms. This can be done based on
simulated data that can always be made \emph{in-task} for the LSP (centered, trend-free and
so on). For that goal we may adopt the literal form of the LSP with its empty $\mathcal
M_1$ in what follows below.

\subsubsection{Frequentist LS periodogram}
Given the comparison models $\mathcal M_{1,2}$ as specified above, the frequentist
likelihood-ratio statistic $R_{\rm F}$ is basically reduced to the chi-square test, just
like the LSP, so the following relationship holds:
\begin{equation}
R_{\rm F}=e^{z_{\max}}, \quad z_{\max} = \max_{f \in [f_{\rm l},f_{\rm u}]} z(f),
\label{lsRF}
\end{equation}
where $z(f)$ is the LSP as function of frequency, and $[f_{\rm l},f_{\rm u}]$ is the
frequency range being scanned.

As we noticed above, thanks to the parameterless $\mathcal M_1$ all three definitions of
the $\FAP$ appear equal to each other, $\alpha_{\rm plug}=\alpha_{\rm sup}=\alpha_{\rm
avg}$. Denoting all them as just $\alpha_{\rm F}$, its approximation for not too small
$z_{\max}$ is given by \citet{Baluev08a}:
\begin{equation}
\alpha_{\rm F} \simeq W e^{-z_{\max}} \sqrt{z_{\max}}, \quad W=(f_{\rm u}-f_{\rm l}) T_{\rm eff},
\label{lsFAPF}
\end{equation}
where $T_{\rm eff}$ being the effective time series length determined through the variance
of $t_i$ (usually $T_{\rm eff}$ is close to the literal time span). In~(\ref{lsFAPF}) we
may substitute $z_{\rm max} = \log R_{\rm F}$, so that $\alpha_{\rm F}$ can be eventually
expressed as function of $R_{\rm F}$.

\subsubsection{Bayesian LS periodogram}
In general, we need two evidences like~(\ref{evB}) to compute the Bayes factor~(\ref{BF}).
The evidence for the parameterless $\mathcal M_1$ is trivial since it is equal to the
likelihood. The evidence for $\mathcal M_2$ may be computed only numerically in general.
However, in case of the sinusoidal signal an analytic approximation to the corresponding
Bayesian odds ratio was obtained by \citet{Cumming04}, based on the quadratic decomposition
of the likelihood function (aka the Laplace method). This derivation was based on the
uniform priors assumed for the signal period $P=1/f$, amplitude $K$, and phase $\varphi$.
In term of the Bayes factor and using our notation their results can be rewritten as
\begin{equation}
R_{\rm B} \simeq M^{-1} e^{z_{\max}}, \quad M\approx \frac{\Delta P}{\delta P} \frac{\Delta K}{\delta K \delta\varphi},
\label{CummingBF}
\end{equation}
where all quantities of the type $\delta a$ stand for the statistical uncertainty of a
parameter $a$, while $\Delta a$ designates the spanning range of its prior. Because these
uncertainties were not explicitly expressed, the formula~(\ref{CummingBF}) is not yet final
in our context.

We therefore transformed~(\ref{CummingBF}) further, by expressing the necessary
$\delta$-uncertainties from the Fisher information matrix of the task. We also tried to
generalize~(\ref{CummingBF}) to arbitrary priors $p_K$ (for the amplitude) and $p_f$ (for
the frequency), but assuming they vary slowly in comparison with the likelihood (and so the
priors can be approximated by constants inside the evidence integrals). The final result
reads like:
\begin{equation}
R_{\rm B} \simeq \frac{e^{z_{\max}}}{z_{\max} \sqrt I}\, p_K\left(2\sqrt{\frac{z_{\max}}{I}}\right) \frac{p_f(f_{\max})}{T_{\rm eff}}, \quad I = \sum_{i=1}^N \frac{1}{\sigma_i^2},
\label{lsBF}
\end{equation}
where $z_{\max}$ and $f_{\max}$ characterize the tallest periodogram peak and its position.

The essential assumptions for the approximation~(\ref{lsBF}) are (i) large $z_{\max}$,
because small $z$ invalidates the Laplace approximation, (ii) the second tall peak
$(z_{\max}',f_{\max}')$, as well as all further periodogram peaks, may generate only much
smaller values of~(\ref{lsBF}). The second limitation appears because if $p_f(f)$ is
nonuniform then the absolute maximum $z_{\max}$ may be penalized by small $p_f(f_{\max})$
in favor of some secondary peak. Whether or not it is important, depends on the particular
$p_f$, but this limitation can be removed completely if $p_f$ is constant:
\begin{equation}
\left. R_{\rm B} \right|_{p_f=\const} \simeq \frac{e^{z_{\max}}}{W z_{\max} \sqrt I}\, p_K\left(2\sqrt{\frac{z_{\max}}{I}}\right).
\label{lsBF2}
\end{equation}

We can see from~(\ref{lsBF2}) that $R_{\rm B}$ is now rendered as a deterministic function
of $z_{\max}$, and since by~(\ref{lsRF}) $R_{\rm F}$ is also a function of $z_{\max}$ there
is a one-to-one mapping between $R_{\rm F}$ and $R_{\rm B}$. Although we have not yet
characterize the type-II mistakes rate (the $\beta$ function), it is now clear that
$\alpha\beta$ diagrams for the $R_{\rm F}$- and $R_{\rm B}$-based tests must be nearly
identical. The quantities $R_{\rm F}$ and $R_{\rm B}$ are simply two different
parametrizations of (nearly) the same $\alpha\beta$ curve (Fig.~\ref{fig_errcomp}, third
panel). This allows us to conclude, even without resorting to any simulations, that
Bayesian and frequentist LSP-based signal detection are practically equivalent \emph{if the
frequency prior is set to be uniform}.

Remarkably, this conclusion is independent of the amplitude prior $p_K$, but matters can be
different if $p_f$ is nonuniform. In this case it follows from~(\ref{lsBF}) that $R_{\rm
B}$ also depends on $f_{\max}$ which is random. This adds some random spread that turns a
deterministic relationship between $R_{\rm F}$ and $R_{\rm B}$ into just a statistical
correlation. Effect of this type may introduce, in theory, some systematic divergence in
the corresponding $\alpha\beta$ curves, which can be characterized by simulations.

\section{Simulations layout}
\label{sec_layout}
\subsection{Time series and priors}
We consider two types of simulated data to be analysed. The first one corresponds to an
even time series with timings $t_k=k$, $k=1,2,\ldots,N$. To track possible dependency on
data amount, we consider three such time series for $N=10,30,100$. Their Nyquist frequency
is always $f_{\rm Ny}=\frac{1}{2}$. The second case is an uneven time series with $t_k$
sampled randomly (following uniform distribution) in the range $[0,N]$. This was considered
for just a single $N=100$. The Nyquist frequency is formally undefined for uneven data.

Notice that period search of uneven data usually appears in a context of the aliasing
problem and distinguishing between different alternative periods
\citep{Dawson10,Baluev12,Hara17}. However, this task stands relatively aside from our
primary topic here (signal detection instead of resolving ambiguities). Our goal in the
context of uneven data is to test another effect, namely the possibility to increase the
frequency range much above the Nyquist limit. Moreover, possible aliasing is actually a
nuisance factor here. This is why we constructed our uneven data so that to avoid any
internal periodic patterns (regular gaps/clumps, etc).

We have three parameters of the putative sinusoidal signal: frequency $f$, amplitude $K$,
and phase $\varphi$. For them, we should specify a joint 3D prior to be used (i) in
computation of the evidence~(\ref{evB}) and of the Bayes factor (\ref{BF}), a
``subjective'' prior, and (ii) in the assessment of type-II mistakes rate $\beta_{\rm
avg}$. In the latter case, the prior serves to model the physical population of objects
that generate sinusoidal signals, so it is an ``objective'' prior. On the best, these two
priors should match each other. However, cases of mismatched priors are also possible, if
our prior knowledge appeared either inaccurate or entirely subjective, so it does not agree
with the physical population. We thereby consider all pairwise combinations of all possible
3D priors.

However, we still need to specify the pool of individual 1D priors characterizing each
signal parameter. For the phase $\varphi$ we always assume uniform prior in the range
$[0,2\pi]$, or $p_\varphi(\varphi)=\frac{1}{2\pi}$. For the amplitude $K$ we assume one of
the following Pareto priors:
\begin{equation}
p_K(K) \propto \left\{ \begin{array}{l} K^0, \nonumber\\ \frac{1}{K}, \nonumber\\ \frac{1}{K^2}, \end{array} \right. \quad K \in \left[\frac{1}{K_{\rm m}},K_{\rm m}\right],\quad K_{\rm m}=10.
\end{equation}
The first one is simply uniform in $K$, while the other two are uniform in $\log K$ and
$\frac{1}{K}$, respectively.

Concerning the frequency, it appears more tricky. The first prior is uniform:
\begin{equation}
p_f(f) = \frac{1}{f_{\rm u}-f_{\rm l}}, \quad f\in [f_{\rm l}, f_{\rm u}],
\end{equation}
where $f_{\rm l}$ is always zero, and $f_{\rm u}$ depends on the time series used in
simulation. For even time series we use the full Nyquist range, so $f_{\rm u} = f_{\rm Ny}
= \frac{1}{2}$. Larger frequency range is meaningless in this context, because any periodic
signal beyond $f_{\rm Ny}$ is equivalent to some alias frequency from the main range, so it
will be naturally interpreted as such. However, uneven data do not formally have a Nyquist
frequency, and allow for much higher-frequency signals to be detected. In this case we
assume a ten times wider frequency range with $f_{\rm u} = 5$.

We also need a nonuniform frequency prior as an alternative, also different for the even
and uneven data. The even case is again special because of the aliasing. Let \emph{physical}
signals follow some populational distribution $p_f^{\rm phys}(f)$, over an unconstrained
$f$-range. But then signals with frequencies $f$ and $f' = f + 2 k f_{\rm Ny}$, for any
integer $k$, are indistinguishable on even data, so the distribution of \emph{observable}
signals becomes:
\begin{equation}
p_f^{\rm obs}(f) = 2\sum_{k=-\infty}^\infty p_f^{\rm phys}(f + 2 k f_{\rm Ny}), \quad f \in [0,f_{\rm Ny}].
\label{wrap}
\end{equation}
This is called the wrapped distribution, and the wrapping usually makes the distribution
more uniform. This means that while we may assume any $p_f^{\rm phys}$, we cannot assume an
arbitrary function for $p_f^{\rm obs}$.

A few wrapped distributions were investigated quite well, in particular they have an
explicitly expressed p.d.f. and c.d.f. We opted to use the wrapped Cauchy distribution, or
WC \citep{MardiaJupp99}. As follows from the title, it originates from the Cauchy physical
distribution:
\begin{equation}
C(f_0):\quad p_f^{\rm phys}(f) = \frac{f_0}{\pi} \frac{1}{f_0^2+f^2}, \quad f\in \mathbb R.
\label{Cauchy}
\end{equation}
A brief review of its properties is given in Appendix~\ref{sec_WC}. In simulations below we
assume $f_0=0.1$, designating this wrapped distribution as ${\rm WC}(0.1)$. This is a
rather nonuniform function, with $\max/\min$ ratio of the p.d.f. about $\sim 10$.

In case of uneven data the wrapping effect does not appear, so we adopt the usual Cauchy
distribution for $p_f$, though cut its tails to fill in the required frequency range (with
$f_{\rm u}=5$). We also set $f_0=1$ now to have roughly the same $\max/\min$ p.d.f. ratio
as for ${\rm WC}(0.1)$.

Notice that Cauchy distribution implies a quadratic probability decrease in the tails,
$p_f(f)\sim \frac{1}{f^2}$, which scales to a constant period distribution (for periods
below $\sim \frac{1}{f_0}$). This constant-period prior was only modified to avoid a p.d.f.
singularity at $f=0$.

\subsection{Simulations scheme}
Let us count our to-do work. We have $1$ phase prior, $3$ amplitude priors, and $2$
frequency priors. Together they can be combined into $6$ variants of the 3D prior for
$(f,K,\varphi)$. We treat separately the ``subjective'' prior used in Bayesian computations
and the ``objective'' prior describing the physical population, so we have $36$ pairwise
combinations. Finally, we have $4$ time series to analyse. This counts, in total, to $144$
simulated data-analysis tasks. For each of them we intend to generate two Monte Carlo
samples by simulating $\bmath x$ as (i) just a WGN vector (model $\mathcal M_1$), and (ii)
WGN+signal (model $\mathcal M_2$). The first simulation sequence can be used to construct
the $\alpha(R_{\rm B})$ curve (through the empiric distribution function of $R_{\rm B}$).
It can be done for just $24$ (instead of $144$) tasks, because $\mathcal M_1$ assumes there
is no physical objects, hence no ``objective'' prior. The second simulation is needed for
$\beta_{\rm avg}(R_{\rm B})$, and it should be done for all $144$ tasks. All this finally
counts to $168$ Monte Carlo samples. And each sample contains $10^4$ trials that involve a
single computation of $R_{\rm B}$ and $R_{\rm F}$ (we always compute them jointly as a
pair).

Now let us provide some details how we can simulate the $\alpha\beta$ diagram. Per each
task, Monte Carlo simulations yield two random samples of our test statistic $\tau$. The
first sample $\{\tau_i^{\rm I}\}_{i=1}^{N_{\rm I}}$ is for model $\mathcal M_1$, and the
second one $\{\tau_i^{\rm II}\}_{i=1}^{N_{\rm II}}$ is for model $\mathcal M_2$. The
$\alpha$ and $\beta$ rates can be formally estimated through empiric distributions of
these samples:
\begin{eqnarray}
\hat\alpha(\tau) &=& 1-\frac{1}{N_{\rm I}} \sum_{i=1}^{N_{\rm I}} H(\tau-\tau_i^{\rm I}), \nonumber\\
\hat\beta(\tau) = \hat\beta_{\rm avg}(\tau) &=& \frac{1}{N_{\rm II}} \sum_{i=1}^{N_{\rm II}} H(\tau-\tau_i^{\rm II}).
\label{abest}
\end{eqnarray}
where $H(x)$ is Heaviside step function. Then, from~(\ref{Iab}), we obtain
\begin{eqnarray}
\hat I_{\alpha\beta} &=& - \int\limits_0^1 \hat\beta(\tau) \hat\alpha'(\tau) d\tau =\nonumber\\
&=& \frac{1}{N_{\rm I}N_{\rm II}} \sum_{i=1}^{N_{\rm II}} \sum_{j=1}^{N_{\rm I}} \int\limits_0^1 H(\tau-\tau_i^{\rm II}) \delta(\tau-\tau_j^{\rm I}) d\tau =\nonumber\\
&=& \frac{1}{N_{\rm I}N_{\rm II}} \sum_{i=1}^{N_{\rm II}} \sum_{j=1}^{N_{\rm I}} H(\tau_j^{\rm I}-\tau_i^{\rm II}),
\label{Iabest}
\end{eqnarray}
so to estimate $I_{\alpha\beta}$ we should count the occurrences of $\tau_j^{\rm
I}>\tau_i^{\rm II}$, for all possible index pairs $(i,j)$.

Since~(\ref{abest}) and~(\ref{Iabest}) are only estimates, they have uncertainties owed to
the Monte Carlo method. In particular, the variance of $\hat I_{\alpha\beta}$ is important
in what follows below. It is derived in Appendix~\ref{sec_vIab}, and reads
\begin{equation}
\disp \hat I_{\alpha\beta} \simeq \frac{1}{N_{\rm II}} \left( \int\limits_0^1 \alpha^2 d\beta - I_{\alpha\beta}^2 \right) + \frac{1}{N_{\rm I}} \left(\int\limits_0^1 \beta^2 d\alpha - I_{\alpha\beta}^2 \right),
\label{vIab}
\end{equation}
where additional integrals (those that involve squared $\alpha,\beta$) can be estimated by
analogy with~(\ref{Iabest}).

\subsection{Computing Bayesian evidence}
To compute $R_{\rm F}$ we only need to numerically maximize the LSP. This is a relatively
fast operation, but the Bayes factor $R_{\rm B}$ appears more challenging, because its
computation involves Monte Carlo simulations in itself.

Multiple approaches are available to compute the evidence integral~(\ref{evB}), see e.g.
\citet{Nelson20} for some practical list. One popular in astronomy method is the nested
sampling algorithm \citep{Skilling04} and its improved versions like {\sc MultiNest}
\citep{Feroz09} or {\sc DNest4} \citep{Brewer16}. Yet another branch is based on the use of
the posterior MCMC sampling output, which is faster to generate
\citep{Weinberg12}. Speaking roughly, these methods rely on the harmonic mean estimator of
the likelihood, but modified so that to get rid of its inconsistency (infinite variance).

However, in the context of our study too complicated tools revealed certain disadvantages,
because the evidence should be computed thousand of times for data with very different
$R_{\rm B}$. This assumes no chance to manually check or tune individual computations.
Without that some computations fail, resulting in an invalid or inaccurate evidence
estimate, likely because of unsuitable tuning parameters or because of other algorithm
issues. Although the fraction of such cases was not very large in our tests, they are not
so easy to identify, and their effect on the final results is unclear.

We therefore opted to use the most simple (and hence reliable in our context) method based
on Monte Carlo integration. We draw the sample $\{\btheta_i\}_{i=1}^{N_{\rm MC}}$ from the
prior $\mathcal W(\btheta)$, and then estimate $E_{\rm B}$ by averaging the resulting
likelihood values:
\begin{equation}
\hat E_{\rm B} = \frac{1}{N_{\rm MC}} \sum_{i=1}^{N_{\rm MC}} L(\bmath x | \btheta_i).
\end{equation}
This method appeared satisfactorily fast in our tasks, and it allows for an easy accuracy
control through the variance estimate:
\begin{equation}
\disp \hat E_{\rm B} \simeq \frac{1}{N_{\rm MC} (N_{\rm MC}-1)} \sum_{i=1}^{N_{\rm MC}} \left[ L(\bmath x | \btheta_i) - \hat E_{\rm B} \right]^2.
\end{equation}
We adopted an accuracy-governed approach to compute $E_{\rm B}$. The starting number of
trials $N_{\rm MC}$ was set to $10^5$, and if necessary we continued the simulation until
getting the ratio $\sqrt{\disp \hat E_{\rm B}}/\hat E_{\rm B}$ below $3\times 10^{-3}$.

\section{Simulations results}
\label{sec_results}
We started our simulations from three even time series with $N=10,30,100$, as a more simple
case. However, here we do not have enough space to present all diagrams generated by these
simulations, so we will show only part of them, mostly for $N=100$. The full set of plots
is available in the online-only supplement.

\begin{figure*}
\includegraphics[width=\linewidth]{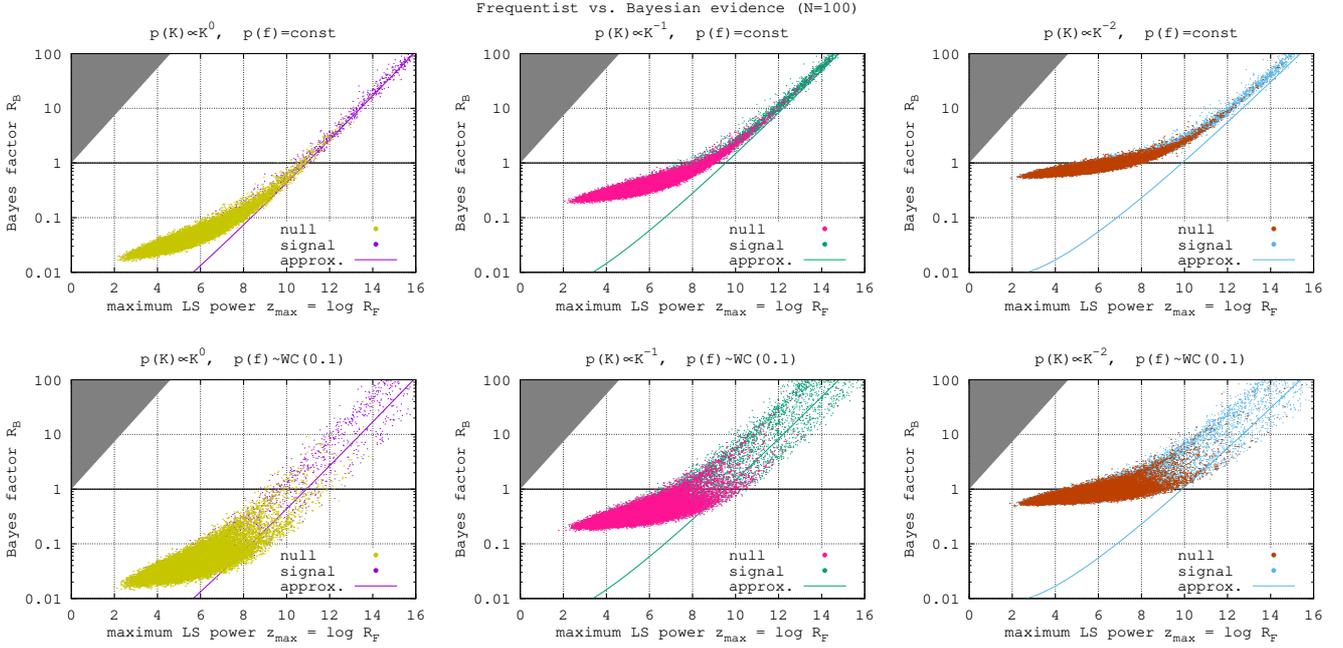}
\caption{Relationship between the frequentist and Bayesian detection metrics
$R_{\rm F}$ and $R_{\rm B}$, assuming $6$ different priors. Each point in a graph
represents a single MC trial, with $10^4$ points for the model $\mathcal M_1$ (``null''
label) and $10^4$ points for $\mathcal M_2$ (``signal''). We also show the
approximation~(\ref{lsBF2}), and the general boundary $R_{\rm B}\leq R_{\rm F}$ that
follows from~(\ref{evB},\ref{evF}). For the ``signal'' part, the population priors coincide
with the ones assumed in $R_{\rm B}$. This plot is for $N=100$ even timings, and analogous
plots for $N=10$ and $N=30$, as well as for $N=100$ random timings, are available in the
online-only supplement.}
\label{fig_RBRF100}
\end{figure*}

First of all, let us verify the relationship~(\ref{lsBF2}) between the $R_{\rm F}$ and
$R_{\rm B}$ metrics. These quantities are plotted against each other in
Fig.~\ref{fig_RBRF100}, for all $6$ available priors (``objective'' and ``subjective''
priors are identical here). Three cases with uniform-frequency priors (top row of plots)
reveal good agreement with~(\ref{lsBF2}), at least for $z_{\max}\gtrsim 10$. Smaller
$z_{\max}$ reveal a systematic deviation from~(\ref{lsBF2}), but some near-deterministic
relationship clearly remains, because the random spread keeps small. This indicates
that~(\ref{lsBF2}) needs a correction in the low-$z$ range, but nevertheless $R_{\rm F}$
and $R_{\rm B}$ remain tightly connected with each other over the entire range. They should
then appear statistically equivalent in terms of their $\alpha\beta$ diagrams. This
conclusion requires uniform $p_f$, and provided that holds for all $p_K$.

We can see significant changes to the picture when the frequency prior becomes nonuniform
(bottom row of Fig.~\ref{fig_RBRF100}). Nearly-deterministic relationship between $R_{\rm
F}$ and $R_{\rm B}$ turns into just a correlation, with a large spread covering about an
order of magnitude. The formula~(\ref{lsBF2}) may still serve as an average prediction to
the Bayes factor, but it can be several times larger or several times smaller than that. In
this case we may possibly expect a statistical difference between these metrics that can be
revealed using the $\alpha\beta$ diagram.

\begin{figure*}
\includegraphics[width=\linewidth]{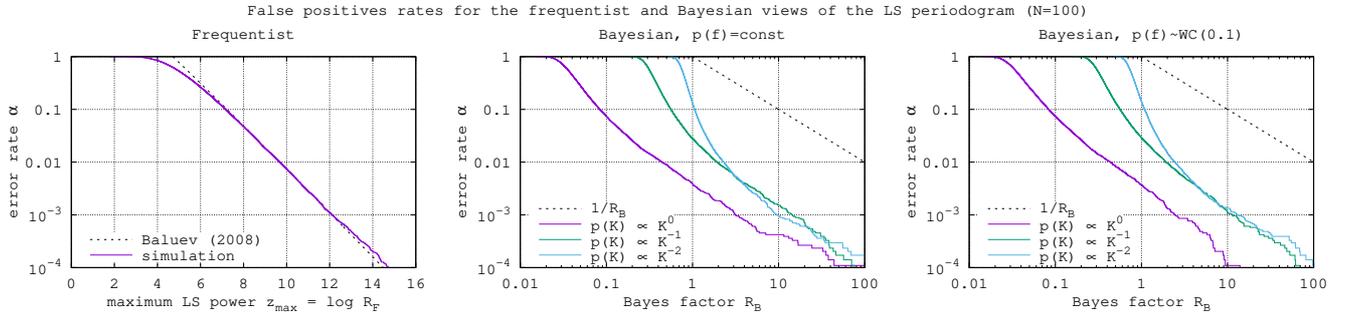}
\caption{Simulated $\alpha$ curves for the frequentist and Bayesian detection metric,
$R_{\rm F}$ and $R_{\rm B}$. While $R_{\rm F}$ does not depend on any prior and is shown as
a single curve, the $R_{\rm B}$ curves are shown for $6$ priors. We also show an analytic
$\FAP$ approximation from \citep{Baluev08a} for $R_{\rm F}$, and the boundary~(\ref{Rineq})
for $R_{\rm B}$. By definition, all curves shown here assume only the $\mathcal M_1$ model,
so population priors do not get involved. This plot is for $N=100$ even timings, and
analogous plots for $N=10$ and $N=30$, as well as for $N=100$ random timings, are available
in the online-only supplement.}
\label{fig_alpha100}
\end{figure*}

But before considering $\alpha\beta$ diagrams, let us investigate how well the Bayes factor
is calibrated against false positives. Recall that there is a common empiric scale for
$R_{\rm B}$ due to \citet{JeffreysTP} that should enable (in theory) its use without
resorting to any mistakes rates. But simultaneously, such a scale can be constructed more
objectively from the $\alpha(R_{\rm B})$ function, on a task-by-task basis. So let us see
how well these two scales agree with each other in the LSP/BLSP framework.

Fig.~\ref{fig_alpha100} demonstrates several $\alpha$ curves for all $6$ priors (still
assuming identical ``subjective'' and ``objective'' ones). Though the curves are rather
different, one common tendency is an unexpectedly small $\alpha$ level for $R_{\rm B}=1$.
This $R_{\rm B}$ level should indicate, based on the empiric scale, statistically equal
models. But for $p_K=\const$ the level $R_{\rm B}=1$ appears actually quite decisive in
favour of $\mathcal M_2$, because it has $\alpha<0.01$. An inconclusive case should occur
for $\alpha\sim 0.5$ that corresponds to $R_{\rm B}<0.1$ (for the same $p_K=\const$).
However, based on the empiric scale such $R_{\rm B}$ would indicate a strong support in
favour of $\mathcal M_1$ over $\mathcal M_2$. Therefore, the empiric scale of $R_{\rm B}$
turns to be very overconservative compared to the $\FAP$ calibration.

Practical consequences of such a miscalibration are likely not that bad, if there is no
mismatch between the ``subjective'' and ``objective'' prior. From further investigation of
Fig.~\ref{fig_alpha100} we can see that overconservative behaviour of $R_{\rm B}$ is tied
to $p_K$ and mainly appears if $p_K$ leans towards larger-amplitude signals. For example,
for $p_K=\const$ possible inaccuracy of the $R_{\rm B}$ threshold is unlikely to affect the
number of detected signals very much because the population is mostly located at much
larger amplitudes. Moreover, for large $R_{\rm B}$ the miscalibration effect gets reduced
regardless of $p_K$, for example $R_{\rm B}\gtrsim 10$ imply more stable $\alpha$ values
that are in a better agreement with the empiric scale.

Still the miscalibration effect is important if the ``subjective'' $p_K$ does not match the
``objective'' one. For example, we could wrongly assume $p_K=\const$ (implying extremely
conservative $R_{\rm B}$) while in actuality the objects are distributed as $p_K\propto
K^{-1}$ or $p_K\propto K^{-2}$. In this case $R_{\rm B}$ that we actually use is
systematically smaller than it would for the correct $p_K$. To make this worse, the $R_{\rm
B}$ threshold is now located where the population is dense. Many detectable signals might
then escape if we disregard the calibration against the $\FAP$. Notice that this
miscalibration effect depends on $p_K$, but does not seem to depend (significantly) on
$p_f$.

From now on, we assume that Bayes factor is always properly calibrated against the $\FAP$.
Under that we mean that the function $\alpha(R_{\rm B})$ is used to treat whether the
observed $R_{\rm B}$ is significant or not. So the results below do not include the
miscalibration effect. This makes the analysis rather similar to the frequentist detection,
however in the Bayesian case $\alpha(R_{\rm B})$ also depends on the adopted (subjective)
prior. It is also possible to include the $\beta$ rate in such a calibration, if necessary,
but this function depends on the subjective and objective priors both.

\begin{figure*}
\includegraphics[width=\linewidth]{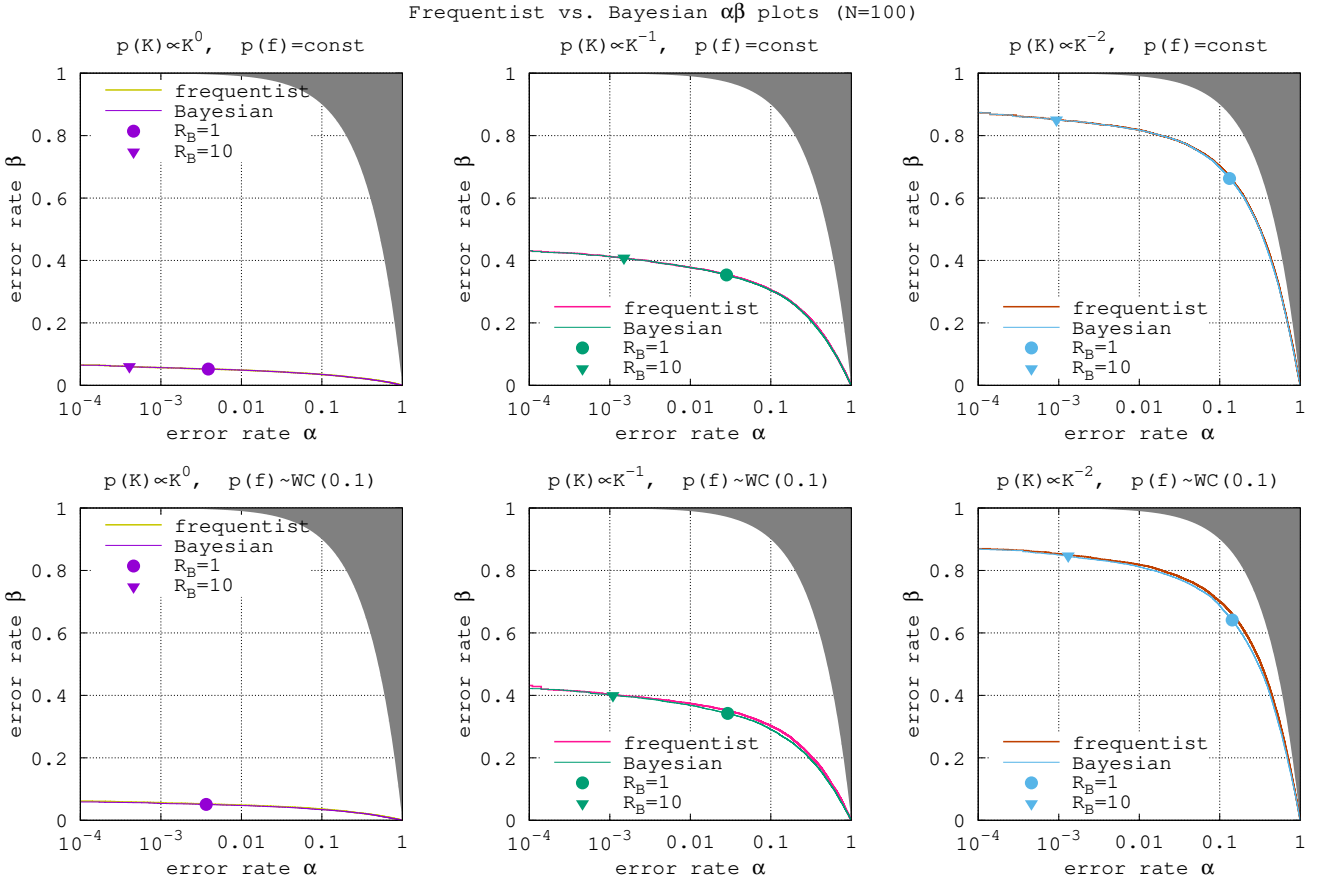}
\caption{The $\alpha\beta$ diagrams intercomparing the frequentist and Bayesian detection
metrics, for $6$ different priors, and for even timings with $N=100$. Each plot shows two
simulated diagrams for $R_{\rm F}$ and $R_{\rm B}$ that always appear to nearly coincide.
The positions of $R_{\rm B}=1$ and  $R_{\rm B}=10$ are also shown for a reference (they
depend on the prior). The population priors coincide with the ones used in $R_{\rm B}$.
Analogous plots for all $36$ combinations of priors, including mismatch cases, and for all
time series, are available in the online-only supplement.}
\label{fig_ab100}
\end{figure*}

Now let us proceed to our final goal, investigation of $\alpha\beta$ diagrams for the
$R_{\rm F}$- and $R_{\rm B}$-based detection. In Fig.~\ref{fig_ab100} they are shown for
all $6$ cases when the ``objective'' and ``subjective'' priors match each other. In each
diagram the frequentist and Bayesian curves nearly coincide, implying that these metrics
are practically equivalent. Such a behaviour was expected for the case of uniform $p_f$,
but for the nonuniform $p_f$ it surprisingly remains visually identical. Even though
individual values of $R_{\rm B}$ have shown large spread because of nonconstant $p_f$, they
generate nearly the same $\alpha\beta$ curve. Hence, there is no practical difference
between Bayesian and frequentist period search, at least in model tasks we considered.

Entirely analogous picture appears for all $N$, and also in all cases with mismatched
priors in all $36$ pairwise combinations (see online-only figures). It does not matter what
the prior we select from our pool, Bayesian analysis always reveals practically the
same $\alpha\beta$ diagram as the frequentist one. In terms of examples shown in
Fig.~\ref{fig_errcomp}, all our simulations were close to the third panel scheme.

\begin{figure}
\includegraphics[width=84mm]{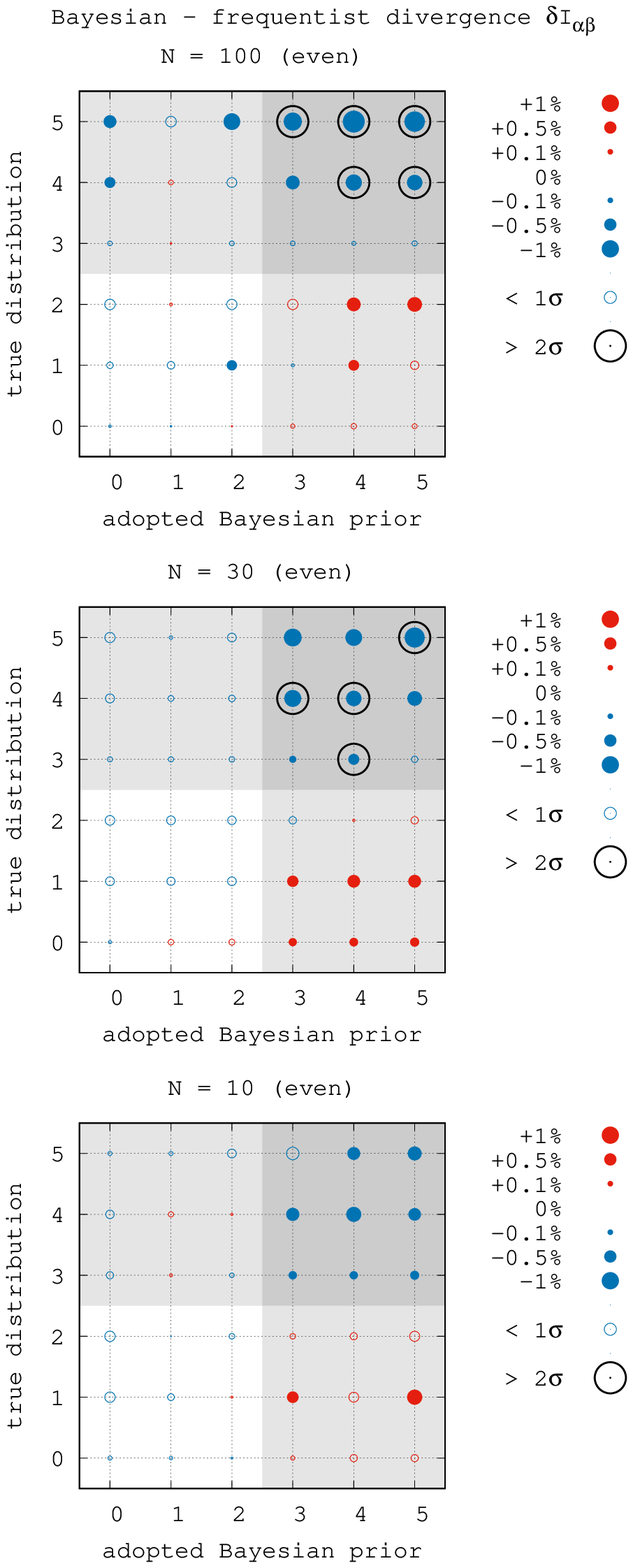}
\caption{Estimated efficiency differences
$\delta I_{\alpha\beta}$ between Bayesian and frequentist signal detection, for all $36$
combinations of priors and $3$ even time series. Coordinates in each plot are integer
indices enumerating the prior (same sequence as for plots in Fig.~\ref{fig_RBRF100}).
Quarters that involve one or both nonuniform frequency priors are additionally highlighted
by gray/darker-gray background. Blue points indicate advantage of Bayesian detection, while
red ones indicate advantage of the frequentist one. The area of each point is proportional
to $|\delta I_{\alpha\beta}|$, but untrustable points that have a greater-than-value
uncertainty (significance below one-sigma) are shown empty. Points with uncertainty smaller
than half of the value (significance above two-sigma) are additionally encircled.}
\label{fig_IabE}
\end{figure}

To develop these findings further, we computed Monte Carlo estimates for all
$I_{\alpha\beta}$, and the associated differences $\delta I_{\alpha\beta}$ between the
$R_{\rm B}$- and $R_{\rm F}$-based detection. These differences are plotted in
Fig.~\ref{fig_IabE} for all $36$ combinations of priors. The values of $\delta
I_{\alpha\beta}$ are demonstrated through sizes of points. Majority of values appeared
consistent with zero (within the uncertainty). A sound (above two sigma) advantage of the
Bayesian analysis over the frequentist one is revealed in just a few cases, whenever the
adopted \emph{and} actual prior involve nonuniform $p_f$. In half cases of mismatched
priors (assuming nonuniform $p_f$ while the true one is uniform) we see an average
advantage of the frequentist approach, though this looks a bit less reliable given the
uncertainties. In case of an opposite mismatch (true $p_f$ is nonuniform but assuming it is
uniform) our $\delta I_{\alpha\beta}$ looks mostly consistent with zero.

We hoped that the case of $N=10$ would reveal the biggest Bayesian/frequentist difference,
because the role of the prior increases for smaller data, but in actuality this case
revealed even smaller $\delta I_{\alpha\beta}$.

\begin{figure*}
\includegraphics[width=\linewidth]{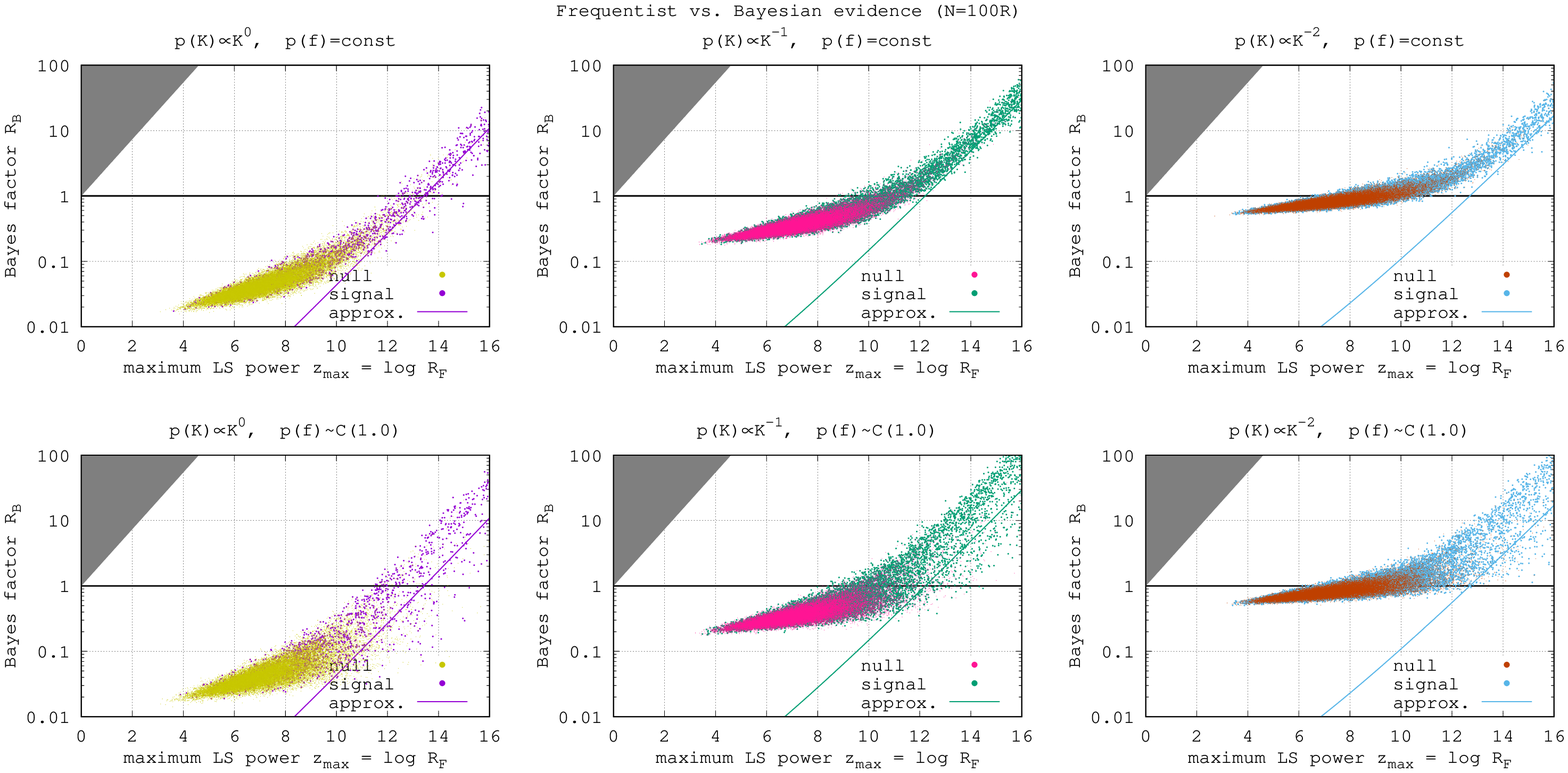}
\caption{Same as Fig.~\ref{fig_RBRF100}, but for $N=100$ \emph{randomly} spaced timings,
and with frequency range increased by the factor of ten above $f_{\rm Ny}$.}
\label{fig_RBRF100R}
\end{figure*}

Finally, we investigated the case of randomly spaced time series. We reiterate once more
that the main reason for that was because random timings allow to consider much wider
frequency range than the Nyquist one. In this test case we had $N=100$ and the maximum
frequency ten times the previous limit $f_{\rm Ny}=\frac{1}{2}$. This corresponds to the
factor $W$ in~(\ref{lsFAPF}) about $\simeq 500$. Consequently, we have roughly $500$ peaks
in a periodogram, so this is a rather large frequency range. In Fig.~\ref{fig_RBRF100R} we
show the $R_{\rm F}$ vs. $R_{\rm B}$ diagram which appears generally similar to
Fig.~\ref{fig_RBRF100} and reveals similar relationships (nearly deterministic one for
uniform $p_f$ and only correlational one for non-uniform $p_f$).

\begin{figure}
\includegraphics[width=84mm]{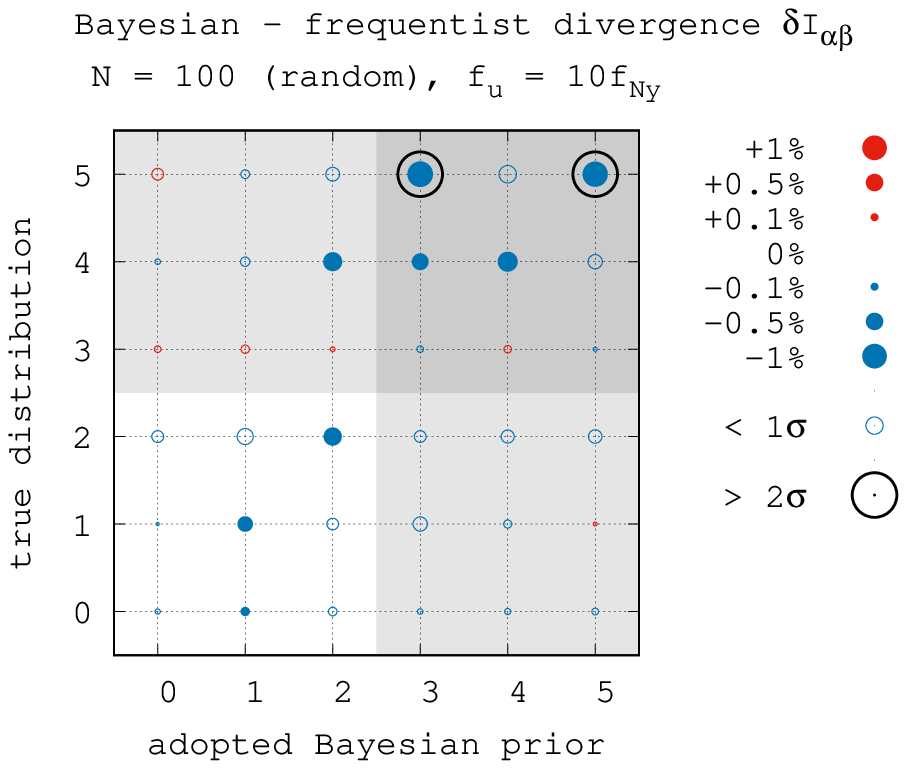}
\caption{Same as plots of Fig.~\ref{fig_IabE}, but for $N=100$ \emph{randomly} spaced timings,
and with frequency range increased by the factor of ten above $f_{\rm Ny}$.}
\label{fig_IabR}
\end{figure}

The simulated $\alpha\beta$ diagrams (see online supplement) again appear nearly identical
for $R_{\rm F}$ and $R_{\rm B}$. The associated estimates of $\delta I_{\alpha\beta}$ are
shown in Fig.~\ref{fig_IabR}, and they reveal only two cases above two-sigma. In general,
these simulations appear quite similar to those for the even data.

In all the cases that we considered, absolute values of $\delta I_{\alpha\beta}$ remain
below $0.02$, and usually below $0.01$. This is very formal and practically negligible
difference, even if measurable. However, can this difference be boosted to a higher level,
for example by tuning the nonuniform $p_f$ distribution?

First of all, it is necessary to derive, at least roughly, what characteristic of $p_f$ may
define the $\alpha\beta$ curve. The frequentist test does not depend on $p_f$ at all, but
Bayesian one does. This effect can be roughly assessed by intercomparing the Bayes factor
approximations~(\ref{lsBF}) and~(\ref{lsBF2}). We can obtain a relationship
\begin{equation}
\log R_{\rm B} \simeq \left. \log R_{\rm B}\right|_{p_f=\const} + \log \left[ f_{\rm u}\, p_f(f_{\max}) \right],
\label{BFrel}
\end{equation}
where $f_{\max}$ being a random quantity, namely the frequency of the maximum periodogram
peak. The distribution of $f_{\max}$ is not necessarily the same as $p_f$ and is also
different for the null and alternative model. If $\mathcal M_1$ is true then there is no
signal in the data and $f_{\max}$ is uniformly distributed in the range $[0,f_{\rm u}]$,
since this is how noisy peaks are distributed in the LSP. If $\mathcal M_2$ is true, the
maximum peak becomes correlated with the existing signal. In the ultimate case, when the
signal amplitude is large, $f_{\max}$ becomes nearly certainly equal to the signal
frequency. If the latter case was always true, the distribution of $f_{\max}$ should be the
same as $p_f$, but in actuality it should be somewhere between the uniform one and the
$p_f$. Still we can adopt $p_f$ as an ultimately nonuniform limit.

From~(\ref{BFrel}) it follows that the first-order statistical effect imposed by the
nonuniform $p_f$ is the bias of $\log R_{\rm B}$:
\begin{equation}
b = \expect \left[ \log R_{\rm B} - \left. \log R_{\rm B}\right|_{p_f=\const}\right] \simeq \log f_{\rm u} + \expect \log p_f(f_{\max}).
\end{equation}
In case of model $\mathcal M_1$, this bias reads
\begin{equation}
b_{\rm I} \simeq \log f_{\rm u} + \frac{1}{f_{\rm u}} \int\limits_0^{f_{\rm u}} \log p_f(f_{\max}) df_{\max},
\end{equation}
and by setting $p_f$ to the Cauchy distribution~(\ref{Cauchy}) we obtain $b_{\rm I} \sim
\const - \log\frac{f_{\rm u}}{f_0}$ (this assumes $f_{\rm u}\gg f_0$). For the model
$\mathcal M_2$, and assuming the $p_f$ distribution for $f_{\max}$, the bias can be
expressed through the entropy $\mathcal H(p_f)$:
\begin{equation}
b_{\rm II} \simeq \log f_{\rm u} + \int\limits_0^{f_{\rm u}} \log p_f(f_{\max}) p_f(f_{\max}) df_{\max} = \log f_{\rm u} - \mathcal H(p_f),
\end{equation}
For the Cauchy distribution taken as $p_f$, we obtain $b_{\rm II} \sim \const +
\log\frac{f_{\rm u}}{f_0}$ (again for $f_{\rm u}\gg f_0$).

We can see that our primary quantity that determines statistical effect of $p_f$ is the
ratio $\frac{f_{\rm u}}{f_0}$. This same quantity should likely define the first-order
effect on the $\alpha\beta$ diagram. However, from~(\ref{lsFAPF}) and~(\ref{lsBF2}) it
follows that the dependency of $\alpha$ rate on $R_{\rm B}$ is roughly logarithmic, so the
effect should be roughly proportional to $\log\frac{f_{\rm u}}{f_0}$. This logarithmic
dependency explains why our simulations revealed so small difference between the
frequentist and Bayesian $\alpha\beta$ diagrams: we had $\frac{f_{\rm u}}{f_0}=5$, and its
logarithm is even a smaller number.

Now it appears not so easy to increase $\delta I_{\alpha\beta}$ by a remarkable amount. For
example, to raise it above $\sim 0.1$ (still a modest level), we should increase
$\log\frac{f_{\rm u}}{f_0}$ by an order of magnitude, and this scales to $\frac{f_{\rm
u}}{f_0} \sim 10^7$. Such an extreme prior does not seem very realistic from the practical
point of view. At least, we should have a sound physical basis justifying so nonuniform
frequency distribution (to justify, for example, why we set tail p.d.f. levels to $\sim
10^{-14}$ instead of just zero). It seems that for a majority of practically reasonable
frequency distributions $\delta I_{\alpha\beta}$ would likely remain about a few per cent,
at most.

\section{Discussion}
\label{sec_discuss}
Our main conclusion is that Bayesian and frequentist analyses reveal nearly equivalent
efficiency in the task of sinusoidal signal detection. In case of the uniform frequency
prior this follows explicitly from analytic approximations that express Bayes factor
through the LSP. Cases involving nonuniform frequency prior reveal sometimes a formal
advantage of Bayesian detection, but this may turn opposite whenever the adopted
frequency prior does not match the actual distribution of periodic signals (and physical
objects behind them). While the magnitude of Bayesian--frequentist difference $\delta
I_{\alpha\beta}$ does depend on how much our frequency distribution is nonuniform, this
dependence is shallow. It does not seem there is an easy way to achieve a large Bayesian
advantage with a reasonable frequency prior.

Moreover, Bayes factor of this task may be rather poorly calibrated against the false alarm
probability. For example, $R_{\rm B}=1$ may correspond to a situation of a clearly
detectable signal (owing to low $\FAP$), rather than to a 50/50 case, as might be expected.
In other words, Bayes factor may appear overconservative without the
calibration.\footnote{Though we must admit it is perhaps better than to be
underconservative, since we do not increase the rate of false positives.} This effect
depends of the signal's amplitude prior, for example it is larger if this prior is biased
towards relatively high-amplitude signals. However, the amplitude prior does not seem to
affect detection efficiency much. The miscalibration effect becomes also more important in
case of a mismatched prior.

Summarizing these results, Bayesian analysis does not appear very advantageous for the
periodic signal detection, at least for the LSP models we were restricted to. In this task
Bayes factor offers practically the same statistical efficiency as the frequentist
periodogram, but requires more computing resources. Besides, additional issues appear
because of miscalibration against the false positives rate.

In this work we considered only the basic formulation of the period search task within the
Lomb model. More complicated cases may include (i) the use of a nontrivial null model, like
linear or nonlinear trend, (ii) nonsinusoidal periodic signals, (iii) signals built from
multiple components (like multiple sinusoids), (iv) dedicated noise models like Gaussian
processes. Whenever the mathematical complexity (nonlinearity) of our models grows, the
divergence between statistical metrics like $R_{\rm F}$ and $R_{\rm B}$ may grow as well.
For example, in \citep{Baluev14a} it was demonstrated that hidden singularities of the
likelihood function may significantly decrease the efficiency of the maximum-likelihood
metric $E_{\rm F}$ and even to render it useless, because maxima are trapped by
singularities. This is in fact an ultimate example of model nonlinearity. The issue can be
solved by using a regularized model that eliminates singularities from the likelihood.
However, Bayesian metric $E_{\rm B}$ may perform more or less well even without such a
regularization, because it relies on integration, which is less sensitive to singularities.
It is thereby interesting to perform a similar calibration of Bayesian analysis for
such models, but this task is for a separate work.

Yet another issue that fell out of our scope is the issue of aliasing. While we considered
uneven timings above, their only purpose was to increase the maximum frequency above the
Nyquist one. But uneven data are usually considered in the context of aliases, that is
regarding the task of distinguishing between nearly equivalent models that involve
alternative periods. In our study possible aliases would be a nuisance factor, so we tried
to suppress the aliasing by drawing uneven timings from the uniform distribution. However,
aliasing comprises a separate challenge that can be solved either by the frequentist or
Bayesian way, and systematic intercomparison of their efficiency is another topic for a
future research.

\section*{Acknowledgements}
This work was supported by a grant 075-15-2020-780 (N13.1902.21.0039) of the Ministry of
Science and Higher Education of the Russian Federation. I would like to additionally thank
the reviewer, Jo\~ao Faria, for their fruitful comments about the manuscript.

\section*{Data availability}
The data underlying this article are available in the article and in its online
supplementary material.

\bibliographystyle{mnras}
\bibliography{bayes}

\begin{thebibliography}{}
\makeatletter
\relax
\def\mn@urlcharsother{\let\do\@makeother \do\$\do\&\do\#\do\^\do\_\do\%\do\~}
\def\mn@doi{\begingroup\mn@urlcharsother \@ifnextchar [ {\mn@doi@}
  {\mn@doi@[]}}
\def\mn@doi@[#1]#2{\def\@tempa{#1}\ifx\@tempa\@empty \href
  {http://dx.doi.org/#2} {doi:#2}\else \href {http://dx.doi.org/#2} {#1}\fi
  \endgroup}
\def\mn@eprint#1#2{\mn@eprint@#1:#2::\@nil}
\def\mn@eprint@arXiv#1{\href {http://arxiv.org/abs/#1} {{\tt arXiv:#1}}}
\def\mn@eprint@dblp#1{\href {http://dblp.uni-trier.de/rec/bibtex/#1.xml}
  {dblp:#1}}
\def\mn@eprint@#1:#2:#3:#4\@nil{\def\@tempa {#1}\def\@tempb {#2}\def\@tempc
  {#3}\ifx \@tempc \@empty \let \@tempc \@tempb \let \@tempb \@tempa \fi \ifx
  \@tempb \@empty \def\@tempb {arXiv}\fi \@ifundefined
  {mn@eprint@\@tempb}{\@tempb:\@tempc}{\expandafter \expandafter \csname
  mn@eprint@\@tempb\endcsname \expandafter{\@tempc}}}

\bibitem[\protect\citeauthoryear{Baluev}{Baluev}{2008}]{Baluev08a}
Baluev R.~V.,  2008, \mnras, 385, 1279

\bibitem[\protect\citeauthoryear{Baluev}{Baluev}{2009a}]{Baluev08b}
Baluev R.~V.,  2009a, \mnras, 393, 969

\bibitem[\protect\citeauthoryear{Baluev}{Baluev}{2009b}]{Baluev09a}
Baluev R.~V.,  2009b, \mnras, 395, 1541

\bibitem[\protect\citeauthoryear{Baluev}{Baluev}{2012}]{Baluev12}
Baluev R.~V.,  2012, \mnras, 422, 2372

\bibitem[\protect\citeauthoryear{Baluev}{Baluev}{2013a}]{Baluev13a}
Baluev R.~V.,  2013a, \mnras, 429, 2052

\bibitem[\protect\citeauthoryear{Baluev}{Baluev}{2013b}]{Baluev13d}
Baluev R.~V.,  2013b, \mnras, 436, 807

\bibitem[\protect\citeauthoryear{Baluev}{Baluev}{2014}]{Baluev14c}
Baluev R.~V.,  2014, Astrophysics, 57, 434

\bibitem[\protect\citeauthoryear{Baluev}{Baluev}{2015a}]{Baluev14b}
Baluev R.~V.,  2015a, \mnras, 446, 1478

\bibitem[\protect\citeauthoryear{Baluev}{Baluev}{2015b}]{Baluev14a}
Baluev R.~V.,  2015b, \mnras, 446, 1493

\bibitem[\protect\citeauthoryear{Bayarri \& Berger}{Bayarri \&
  Berger}{2000}]{BB00}
Bayarri M.~J.,  Berger J.~O.,  2000, Journal of the American Statistical
  Association, 95, 1127

\bibitem[\protect\citeauthoryear{Bayarri \& Berger}{Bayarri \&
  Berger}{2013}]{Bayes18}
Bayarri M.~J.,  Berger J.~O.,  2013, in \cite{BayesTheoryApp}, Chapt.~18, pp
  361--394, Chapt.~18, pp 361--394

\bibitem[\protect\citeauthoryear{Box}{Box}{1980}]{Box80}
Box G. E.~P.,  1980, J. Roy. Stat. Soc. A, 143, 383

\bibitem[\protect\citeauthoryear{Bretthorst}{Bretthorst}{2001a}]{Bretthorst01a}
Bretthorst G.~L.,  2001a, AIP Conf. Proc., 568, 241

\bibitem[\protect\citeauthoryear{Bretthorst}{Bretthorst}{2001b}]{Bretthorst01b}
Bretthorst G.~L.,  2001b, AIP Conf. Proc., 568, 246

\bibitem[\protect\citeauthoryear{Brewer \& Foreman-Mackey}{Brewer \&
  Foreman-Mackey}{2016}]{Brewer16}
Brewer B.~J.,  Foreman-Mackey D.,  2016, arxiv.org:1606.03757

\bibitem[\protect\citeauthoryear{Cumming}{Cumming}{2004}]{Cumming04}
Cumming A.,  2004, \mnras, 354, 1165

\bibitem[\protect\citeauthoryear{Cumming, Marcy  \& Butler}{Cumming
  et~al.}{1999}]{Cumming99}
Cumming A.,  Marcy G.~W.,   Butler R.~P.,  1999, \apj, 526, 890

\bibitem[\protect\citeauthoryear{Damien, Dellaportas, Polson  \&
  Stephens}{Damien et~al.}{2013}]{BayesTheoryApp}
Damien P.,  Dellaportas P.,  Polson N.~G.,   Stephens D.~A.,  eds, 2013,
  Bayesian Theory and Applications.
Oxford University Press, Oxford

\bibitem[\protect\citeauthoryear{Dawson \& Fabrycky}{Dawson \&
  Fabrycky}{2010}]{Dawson10}
Dawson R.~I.,  Fabrycky D.~C.,  2010, \apj, 722, 937

\bibitem[\protect\citeauthoryear{Draper}{Draper}{2013}]{Bayes20}
Draper D.,  2013, in \cite{BayesTheoryApp}, Chapt.~20, pp 409--431, Chapt.~20,
  pp 409--431

\bibitem[\protect\citeauthoryear{Feroz, Hobson  \& Bridges}{Feroz
  et~al.}{2009}]{Feroz09}
Feroz F.,  Hobson M.~P.,   Bridges M.,  2009, \mnras, 398, 1601

\bibitem[\protect\citeauthoryear{Ferraz-Mello}{Ferraz-Mello}{1981}]{FerrazMello81}
Ferraz-Mello S.,  1981, \aj, 86, 619

\bibitem[\protect\citeauthoryear{Gelman}{Gelman}{2008}]{Gelman08}
Gelman A.,  2008, Bayesian Analysis, 3, 445

\bibitem[\protect\citeauthoryear{Hara, Bou{\'e}, Laskar  \& Correia}{Hara
  et~al.}{2017}]{Hara17}
Hara N.~C.,  Bou{\'e} G.,  Laskar J.,   Correia A. C.~M.,  2017, \mnras, 464,
  1220

\bibitem[\protect\citeauthoryear{Jeffreys}{Jeffreys}{1998}]{JeffreysTP}
Jeffreys S.~H.,  1998, Theory of Probability (The International series of
  monographs on physics) 3rd Edition.
Oxford University Press

\bibitem[\protect\citeauthoryear{Koroluk, Portenko, Skorokhod  \&
  Turbin}{Koroluk et~al.}{1985}]{Koroluk}
Koroluk V.~S.,  Portenko N.~I.,  Skorokhod A.~V.,   Turbin A.~F.,  1985, A
  handbook on the probability theory and mathematical statistics (in Russian).
Naukova dumka, Kyev

\bibitem[\protect\citeauthoryear{Lomb}{Lomb}{1976}]{Lomb76}
Lomb N.~R.,  1976, \apss, 39, 447

\bibitem[\protect\citeauthoryear{Mardia \& Jupp}{Mardia \&
  Jupp}{1999}]{MardiaJupp99}
Mardia K.~V.,  Jupp P.~E.,  1999, Directional Statistics.
John Wiley \& Sons Ltd

\bibitem[\protect\citeauthoryear{Mortier, Faria, Correia, Santerne  \&
  Santos}{Mortier et~al.}{2014}]{Mortier14}
Mortier A.,  Faria J.~P.,  Correia C.~M.,  Santerne A.,   Santos N.~C.,  2014,
  \aap, 573, A101

\bibitem[\protect\citeauthoryear{Nelson et~al.,}{Nelson
  et~al.}{2020}]{Nelson20}
Nelson B.~E.,  et~al., 2020, \aj, 159, 73

\bibitem[\protect\citeauthoryear{Scargle}{Scargle}{1982}]{Scargle82}
Scargle J.~D.,  1982, \apj, 263, 835

\bibitem[\protect\citeauthoryear{Schuster}{Schuster}{1898}]{Schuster1898}
Schuster A.,  1898, Terrestial Magnetism and Atmospheric Electricity, 3, 13

\bibitem[\protect\citeauthoryear{Schwarzenberg-Czerny}{Schwarzenberg-Czerny}{1998}]{SchwCzerny98b}
Schwarzenberg-Czerny A.,  1998, Baltic Astron., 7, 43

\bibitem[\protect\citeauthoryear{Skilling}{Skilling}{2004}]{Skilling04}
Skilling J.,  2004, AIP Conf. Proc., 735, 395

\bibitem[\protect\citeauthoryear{Van{\'\i}{\v{c}}ek}{Van{\'\i}{\v{c}}ek}{1969}]{Vanicek69}
Van{\'\i}{\v{c}}ek P.,  1969, \apss, 4, 387

\bibitem[\protect\citeauthoryear{Weinberg}{Weinberg}{2012}]{Weinberg12}
Weinberg M.~D.,  2012, Bayesian Analysis, 7, 737

\bibitem[\protect\citeauthoryear{Zechmeister \& K{\"u}rster}{Zechmeister \&
  K{\"u}rster}{2009}]{ZechKur09}
Zechmeister M.,  K{\"u}rster M.,  2009, \aap, 496, 577

\makeatother
\end{thebibliography}

\appendix

\section{Deriving inequality for Bayesian prior predictive $\FAP$}
\label{sec_BFAP}
Consider Bayesian $\FAP$ based on the prior predictive average:
\begin{eqnarray}
\alpha_{\rm avg}(R_{\rm B}^*) &=& \int \FAP(R_{\rm B}^* | \btheta_1) \mathcal W_1(\btheta_1) d\btheta_1, \nonumber\\
\FAP(R_{\rm B}^* | \btheta_1) &=& \int\limits_{E_{\rm B}(\bmath x | \mathcal M_2, \mathcal W_2) \geq \atop \phantom{\geq} R_{\rm B}^* E_{\rm B}(\bmath x | \mathcal M_1, \mathcal W_1)} L(\bmath x | \btheta_1, \mathcal M_1) d\bmath x.
\end{eqnarray}
By substituting the second equation into the first one and interchanging integrals, we can
simplify this to
\begin{equation}
\alpha_{\rm avg}(R_{\rm B}^*) = \int\limits_{E_{\rm B}(\bmath x | \mathcal M_2, \mathcal W_2) \geq \atop \phantom{\geq} R_{\rm B}^* E_{\rm B}(\bmath x | \mathcal M_1, \mathcal W_1)}
  E_{\rm B}(\bmath x | \mathcal M_1, \mathcal W_1)\, d\bmath x.
\label{FAPB}
\end{equation}
The $\beta$ measure can be expressed in a similar way:
\begin{equation}
1-\beta_{\rm avg}(R_{\rm B}^*) = \int\limits_{E_{\rm B}(\bmath x | \mathcal M_2, \mathcal W_2) \geq \atop \phantom{\geq} R_{\rm B}^* E_{\rm B}(\bmath x | \mathcal M_1, \mathcal W_1)}
  E_{\rm B}(\bmath x | \mathcal M_2, \mathcal W_2)\, d\bmath x.
\label{TAPB}
\end{equation}
We can see that the integration domains in~(\ref{FAPB}) and~(\ref{TAPB}) are determined
identically. Moreover, the inequality that defines this domain applies to the integrands
within this domain, hence to the integrals themselves. This results in the following
inequality:
\begin{equation}
1-\beta_{\rm avg}(R_{\rm B}^*) \geq R_{\rm B}^* \alpha_{\rm avg}(R_{\rm B}^*).
\end{equation}
Then~(\ref{Rineq}) follows immediately.

\section{The wrapped Cauchy distribution}
\label{sec_WC}
\begin{figure*}
\includegraphics[width=\linewidth]{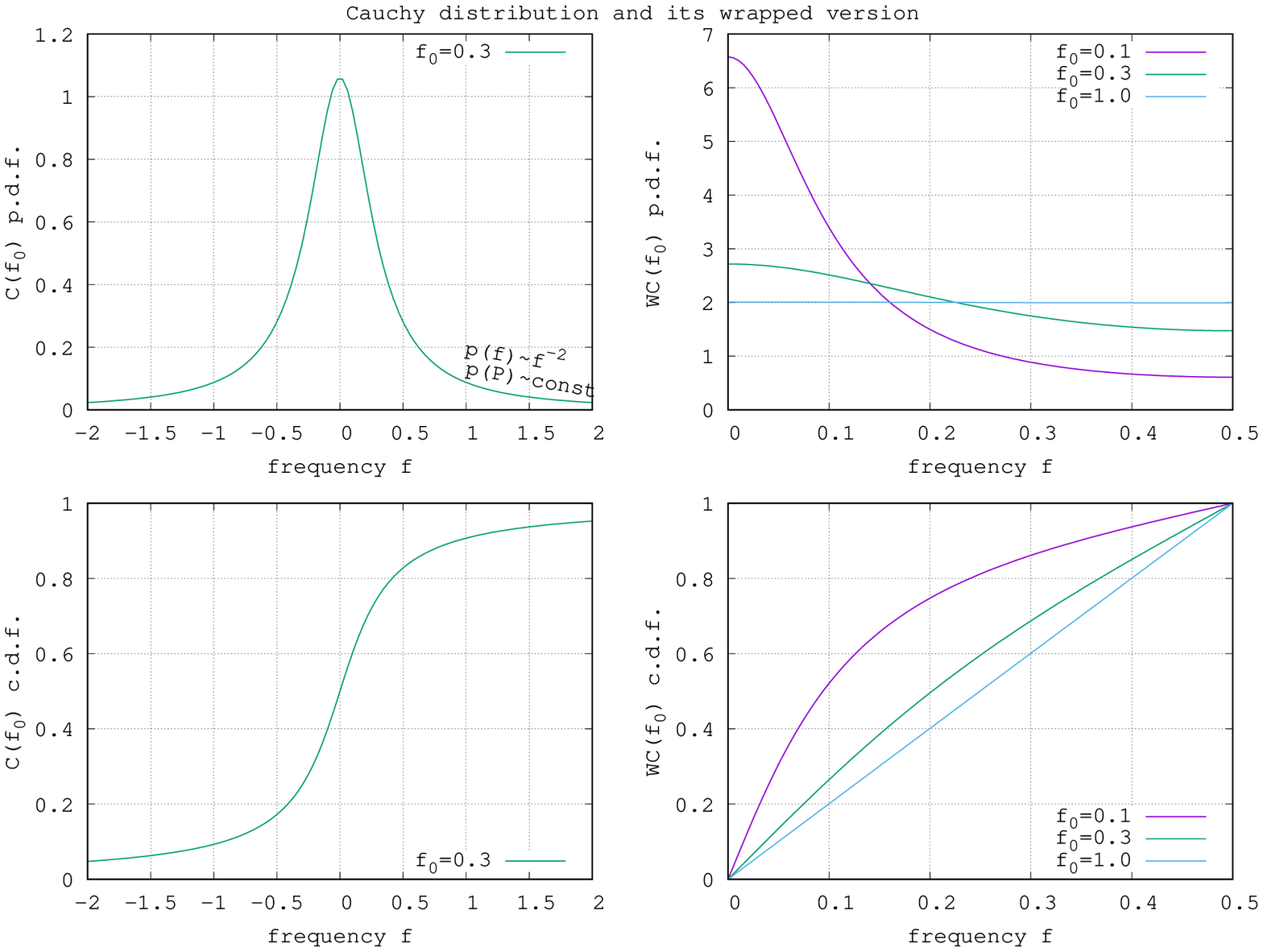}
\caption{A few sample plots of the wrapped Cauchy distribution and of the corresponding
Cauchy distribution it originates from. The Nyquist frequency is $f_{\rm Ny}=\frac{1}{2}$.}
\label{fig_WC}
\end{figure*}

The wrapped Cauchy (WC) distribution is obtained by applying the wrapping
transform~(\ref{wrap}) to the p.d.f.~(\ref{Cauchy}). WC p.d.f. and c.d.f. are known and
both can be expressed by simple analytic formulae \citep[\S~3.5.7]{MardiaJupp99}. In our
normalizations they look like
\begin{eqnarray}
\mathrm{WC}(f_0): & & \nonumber\\
p_f^{\rm obs}(f) &=& \frac{2\sinh(2\pi f_0)}{\cosh(2\pi f_0) - \cos(2\pi f)}, \nonumber\\
P_f^{\rm obs}(f) &=& \frac{1}{\pi}\arccos\frac{\cosh(2\pi f_0) \cos(2\pi f) - 1}{\cosh(2\pi f_0) - \cos(2\pi f)}, \nonumber\\
& & f \in \left[0,f_{\rm Ny}=\frac{1}{2}\right].
\label{WC}
\end{eqnarray}
A few sample plots of this functions are shown in Fig.~\ref{fig_WC}.

The function $P_f^{\rm obs}(f)$ is also easily invertible:
\begin{equation}
f = \frac{1}{2\pi} \arccos\frac{1 + \cosh(2\pi f_0) \cos(\pi P_f^{\rm obs})}{\cosh(2\pi f_0) + \cos(\pi P_f^{\rm obs})}.
\label{WCinv}
\end{equation}
This enables a quick generation of random numbers from the WC distribution, by replacing
$P_f^{\rm obs}$ in~(\ref{WCinv}) with a standard uniform random variate.

\section{Computing the variance of
$I_{\alpha\beta}$ estimate}
\label{sec_vIab}
First of all, from~(\ref{Iabest}) we have
\begin{equation}
\expect \hat I_{\alpha\beta} = \frac{1}{N_{\rm I}N_{\rm II}} \sum_{i=1}^{N_{\rm II}} \sum_{j=1}^{N_{\rm I}} \Pr(\tau_j^{\rm I}>\tau_i^{\rm II}),
\label{EIab}
\end{equation}
because $\expect H(*)$ has the meaning of probability. Now let us express the variance
of~(\ref{Iabest}) by taking its square, then applying the mathematical expectation
operator, and subtracting the squared~(\ref{EIab}):
\begin{eqnarray}
\disp \hat I_{\alpha\beta} &=& \frac{1}{N_{\rm I}^2 N_{\rm II}^2} \sum_{i=1}^{N_{\rm II}} \sum_{j=1}^{N_{\rm I}} \sum_{k=1}^{N_{\rm II}} \sum_{l=1}^{N_{\rm I}} \left\{ \expect\left[H(\tau_j^{\rm I}-\tau_i^{\rm II}) H(\tau_l^{\rm I}-\tau_k^{\rm II})\right] - \right. \nonumber\\
& & \left. - \Pr(\tau_j^{\rm I}>\tau_i^{\rm II}) \Pr(\tau_l^{\rm I}>\tau_k^{\rm II}) \right\}.
\label{DI1}
\end{eqnarray}
Now the behaviour of the common term $T=\expect[H(*)H(*)]$ depends on whether the indices
$(i,j,k,l)$ are all different or some of them coincide. There are four distinct cases:
\begin{enumerate}
\item $i=k$ and $j=l$ implies that $T=\expect H^2(\tau_j^{\rm I}-\tau_i^{\rm II})=\Pr(\tau_j^{\rm I}>\tau_i^{\rm II})$;
\item $i=k$ but $j\neq l$ implies that $T=\expect H[\min(\tau_j^{\rm I},\tau_l^{\rm I})-\tau_i^{\rm II}] = \Pr[\tau_i^{\rm II}<\min(\tau_j^{\rm I},\tau_l^{\rm I})]$;
\item $j=l$ but $i\neq k$ implies that $T=\expect H[\tau_j^{\rm I}-\max(\tau_i^{\rm II},\tau_k^{\rm II})] = \Pr[\tau_j^{\rm I}>\max(\tau_i^{\rm II},\tau_k^{\rm II})]$;
\item $j\neq l$ and $i\neq k$ represents the most frequent case and implies that $T=\Pr(\tau_j^{\rm I}>\tau_i^{\rm II}) \Pr(\tau_l^{\rm I}>\tau_k^{\rm II})$ that cancels with the identical product subtracted in~(\ref{DI1}).
\end{enumerate}
Summing all this up, we have
\begin{eqnarray}
\disp \hat I_{\alpha\beta} &=& \frac{1}{N_{\rm I}^2 N_{\rm II}^2} \sum_{i=1}^{N_{\rm II}} \sum_{j=1}^{N_{\rm I}} \left\{ \Pr(\tau_j^{\rm I}>\tau_i^{\rm II}) - \left[\Pr(\tau_j^{\rm I}>\tau_i^{\rm II})\right]^2 \right\} + \nonumber\\
&+& \frac{1}{N_{\rm I}^2 N_{\rm II}^2} \sum_{i=1}^{N_{\rm II}} \sum_{j=1}^{N_{\rm I}} \sum_{l=1\atop l\neq j}^{N_{\rm I}} \left\{ \Pr[\tau_i^{\rm II}<\min(\tau_j^{\rm I},\tau_l^{\rm I})] - \right. \nonumber\\
& & \left. - \Pr(\tau_j^{\rm I}>\tau_i^{\rm II}) \Pr(\tau_l^{\rm I}>\tau_i^{\rm II}) \right\} + \nonumber\\
&+& \frac{1}{N_{\rm I}^2 N_{\rm II}^2} \sum_{i=1}^{N_{\rm II}} \sum_{j=1}^{N_{\rm I}} \sum_{k=1\atop k\neq i}^{N_{\rm II}} \left\{ \Pr[\tau_j^{\rm I}>\max(\tau_i^{\rm II},\tau_k^{\rm II})] - \right. \nonumber\\
& & \left. - \Pr(\tau_j^{\rm I}>\tau_i^{\rm II}) \Pr(\tau_j^{\rm I}>\tau_k^{\rm II}) \right\}.
\label{DI2}
\end{eqnarray}
Now, let us use two identities $\Pr(\tau^{\rm I}>t) = \expect \alpha(t)$ and $\Pr(\tau^{\rm
II}<t) = \expect \beta(t)$ to transform~(\ref{DI2}) as follows:
\begin{eqnarray}
\disp \hat I_{\alpha\beta} &=& \frac{1}{N_{\rm I}^2 N_{\rm II}^2} \sum_{i=1}^{N_{\rm II}} \sum_{j=1}^{N_{\rm I}} \left\{ \expect\beta(\tau_j^{\rm I}) - \left[\expect\beta(\tau_j^{\rm I})\right]^2 \right\} + \nonumber\\
&+& \frac{1}{N_{\rm I}^2 N_{\rm II}^2} \sum_{i=1}^{N_{\rm II}} \sum_{j=1}^{N_{\rm I}} \sum_{l=1\atop l\neq j}^{N_{\rm I}} \left\{ \expect\beta[\min(\tau_j^{\rm I},\tau_l^{\rm I})] - \right. \nonumber\\
& & \left. - \expect\beta(\tau_j^{\rm I}) \expect\beta(\tau_l^{\rm I}) \right\} + \nonumber\\
&+& \frac{1}{N_{\rm I}^2 N_{\rm II}^2} \sum_{i=1}^{N_{\rm II}} \sum_{j=1}^{N_{\rm I}} \sum_{k=1\atop k\neq i}^{N_{\rm II}} \left\{ \expect\alpha[\max(\tau_i^{\rm II},\tau_k^{\rm II})] - \right. \nonumber\\
& & \left. - \expect\alpha(\tau_i^{\rm II}) \expect\alpha(\tau_k^{\rm II}) \right\}.
\label{DI3}
\end{eqnarray}
Here $\expect\alpha(\tau^{\rm II})=\expect\beta(\tau^{\rm I})=I_{\alpha\beta}$ by
definition~(\ref{Iab}). Also, because $\alpha,\beta$ are monotonic we can replace
$\alpha[\min(x,y)]=\max[\alpha(x),\alpha(y)]$ and
$\beta[\min(x,y)]=\min[\beta(x),\beta(y)]$:
\begin{eqnarray}
\disp \hat I_{\alpha\beta} &=& \frac{1}{N_{\rm I}^2 N_{\rm II}^2} \sum_{i=1}^{N_{\rm II}} \sum_{j=1}^{N_{\rm I}} \left( I_{\alpha\beta} - I_{\alpha\beta}^2 \right) + \nonumber\\
&+& \frac{1}{N_{\rm I}^2 N_{\rm II}^2} \sum_{i=1}^{N_{\rm II}} \sum_{j=1}^{N_{\rm I}} \sum_{l=1\atop l\neq j}^{N_{\rm I}} \left\{ \expect\min[\beta(\tau_j^{\rm I}),\beta(\tau_l^{\rm I})] - I_{\alpha\beta}^2 \right\} + \nonumber\\
&+& \frac{1}{N_{\rm I}^2 N_{\rm II}^2} \sum_{i=1}^{N_{\rm II}} \sum_{j=1}^{N_{\rm I}} \sum_{k=1\atop k\neq i}^{N_{\rm II}} \left\{ \expect\min[\alpha(\tau_i^{\rm II}),\alpha(\tau_k^{\rm II})] - I_{\alpha\beta}^2 \right\}.
\label{DI4}
\end{eqnarray}
Now let us deal with, for example, the following term:
\begin{eqnarray}
\expect\min[\beta(\tau_j^{\rm I}),\beta(\tau_l^{\rm I})] = \int\limits_0^1\int\limits_0^1 \min(\beta_j,\beta_l) d\alpha_j d\alpha_l = \nonumber\\
= \int\limits_0^1 d\alpha_j \left[ \int\limits_0^{\alpha_j} \min(\beta_j,\beta_l) d\alpha_l + \int\limits_{\alpha_j}^1 \min(\beta_j,\beta_l) d\alpha_l \right].
\end{eqnarray}
Since two inner integrals have such limits that either $\alpha_l<\alpha_j$ or
$\alpha_l>\alpha_j$, this implies that $\beta_l>\beta_j$ or $\beta_l<\beta_j$ inside them,
respectively. Therefore, the minima can be replaced with either $\beta_j$ or $\beta_l$, and
\begin{eqnarray}
\expect\min[\beta(\tau_j^{\rm I}),\beta(\tau_l^{\rm I})] = \int\limits_0^1 d\alpha_j \left[ \int\limits_0^{\alpha_j} \beta_j d\alpha_l + \int\limits_{\alpha_j}^1 \beta_l d\alpha_l \right] = \nonumber\\
= \int\limits_0^1 \beta_j d\alpha_j \int\limits_0^{\alpha_j} d\alpha_l + \int\limits_0^1 d\alpha_j \int\limits_{\alpha_j}^1 \beta_l d\alpha_l = \nonumber\\
= \int\limits_0^1 \alpha_j \beta_j d\alpha_j + \int\limits_0^1 d\alpha_l \int\limits_0^{\alpha_l} \beta_l d\alpha_j = 2 \int\limits_0^1 \alpha \beta d\alpha = \int\limits_0^1 \alpha^2 d\beta.
\end{eqnarray}
The term with $\min[\alpha(\tau_i^{\rm II}),\alpha(\tau_k^{\rm II})]$ can be expressed by
analogy and~(\ref{DI4}) finally turns into
\begin{eqnarray}
\disp \hat I_{\alpha\beta} &=& \frac{I_{\alpha\beta} - I_{\alpha\beta}^2}{N_{\rm I} N_{\rm II}} + \nonumber\\
&+& \frac{N_{\rm I}-1}{N_{\rm I} N_{\rm II}} \left( \int\limits_0^1 \alpha^2 d\beta - I_{\alpha\beta}^2 \right) + \nonumber\\
&+& \frac{N_{\rm II}-1}{N_{\rm I} N_{\rm II}} \left(\int\limits_0^1 \beta^2 d\alpha - I_{\alpha\beta}^2 \right).
\label{DI5}
\end{eqnarray}
By neglecting small terms we obtain~(\ref{vIab}).

\section{Online-only material}
\label{sec_OO}
An archive with a complete set of figures in scaleable EPS format. The detailed description
of each figure is given in the Readme file inside the archive.

\bsp

\label{lastpage}

\end{document}